\documentclass[a4paper,12pt]{article}
\usepackage[utf8]{inputenc}
\usepackage{authblk}
\usepackage{graphicx}
\usepackage{caption}
\usepackage{amsmath,amssymb}
\usepackage{pifont}
\usepackage{jcappub}
\usepackage{subcaption}
\usepackage{float}
\pdfoutput=1

\title{Accelerating the search for Axion-Like Particles with machine learning}
\author[a]{Francesca Day}
\author[b]{Sven Krippendorf}
\affiliation[a]{DAMTP, University of Cambridge, Wilberforce Road, Cambridge, United Kingdom}
\affiliation[b]{Arnold Sommerfeld Center for Theoretical Physics, LMU, Theresienstr. 37, 80333 M\"unchen, Germany}

\emailAdd{francesca.day@maths.cam.ac.uk}
\emailAdd{sven.krippendorf@physik.lmu.de}

\abstract{
Machine learning (ML) techniques have been applied with tremendous success in many areas of physics. In this work, we use ML to place bounds on the coupling between photons and axion-like particles (ALPs). This coupling causes ALPs and photons to interconvert in the presence of a background magnetic field. This would lead to modulations in the spectra of point sources shining through the magnetic fields of galaxy clusters. This effect has already been used to place world-leading bounds on the ALP-photon coupling using conventional statistical methods. We train ML classification algorithms on simulated spectra from the {\it Chandra} X-ray telescope for a range of point sources and ALP-photon couplings. We then use the response of these algorithms to the real {\it Chandra} spectra to place bounds on ALP-photon interactions. We obtain bounds at a similar level to those based on other techniques, but find improvements on an individual source basis. We expect such search techniques to become increasingly important for ALP searches with future telescopes that will offer substantially higher energy resolution.
}

\begin{document}
\begin{flushright} LMU-ASC 29/19
\end{flushright}
\maketitle
\section{Introduction}

Axion-like particles (ALPs) are very well motivated extensions of Beyond-The-Standard-Model physics. Utilising the fact that ALPs and photons interconvert in background magnetic fields~\cite{Raffelt:1987im}, there has been a long-standing experimental quest for such particles~\cite{Tanabashi:2018oca}. The ALP-photon Lagrangian is:

\begin{equation}
\mathcal{L} =  \frac{1}{2} \partial_{\mu} a \partial^{\mu} a - \frac{1}{2} m_a^2 a^2 + g_{a \gamma \gamma} a {\bf E} \cdot {\bf B}~,
\end{equation}

\noindent where $a$ is the ALP field, $m_a$ is the ALP mass, $g_{a \gamma \gamma}$ is the coupling between ALPs and photons, ${\bf E}$ is the electric field and ${\bf B}$ is the magnetic field. Linearising the resulting Euler-Lagrange equations for an ALP/photon of frequency $\omega$, we obtain the equation of motion for ALP-photon interconversion in a background magnetic field:

\begin{equation}
\label{conversion}
\left( \omega + \left( \begin{array}{ccc}
\Delta_{\gamma} & 0 & \Delta_{\gamma a x} \\
0 & \Delta_{\gamma} & \Delta_{\gamma a y} \\
\Delta_{\gamma a x}  & \Delta_{\gamma a y} & \Delta_{a} \end{array} \right)
- i \partial_{z} \right) \left( \begin{array}{c}
\mid \gamma_{x} \rangle\\
\mid \gamma_{y} \rangle\\
\mid a \rangle \end{array} \right) = 0~,
\end{equation}
where $\Delta_{\gamma} = \frac{-\omega_{pl}^{2}} {2 \omega}$,  $\Delta_{a} = \frac{-m_{a}^{2}}{\omega}$ and $\Delta_{\gamma a i} = g_{a \gamma \gamma} \frac{B_{i}}{2}$. The effective photon  mass is given by the plasma frequency $\omega_{pl} = \left( 4 \pi \alpha \frac{n_{e}}{m_{e}} \right) ^ {\frac{1}{2}}$, where $\alpha$ is the fine structure constant, $n_e$ is the electron density and $m_e$ is the electron mass. In this work, we will neglect the ALP mass, setting $m_a = 0$. This approximation is valid for ALP masses below the effective photon mass in astrophysical plasmas, $m_a \lesssim 10^{-12} \, {\rm eV}$. Note that in this work we consider generic ALPs, rather than the QCD axion. $m_a$ and $g_{a \gamma \gamma}$ are therefore independent parameters. Equation \ref{conversion} may be solved analytically in certain regimes, but in general requires numerical solution. In either case, we find that the conversion probability is pseudo-sinusoidal in $\frac{1}{\omega}$.

Here we focus on ALP-photon interconversion in the magnetic fields of galaxy clusters. The presence of ultra-light ALPs $(m_{a}\leq 10^{-12}{\rm eV})$) leads to spectral distortions of point sources shining through galaxy clusters at X-ray energies ~\cite{1304.0989}. The search for modulations in the spectra of X-ray sources located in or behind galaxy clusters has lead to world leading bounds on $g_{a \gamma \gamma}$~\cite{1605.01043,1703.07354,1704.05256,1907.05475}. Future X-ray telescopes such as {\it Athena} and  {\it IXPE} will lead to an improvement in these bounds~\cite{1707.00176,1801.10557}.

The search for spectral modulations (reviewed in Section~\ref{sec:problem}) has so far used relatively simple statistical methods and it seems very plausible that search strategies adapted for ALP-like signals might provide higher sensitivity. Machine learning has been used with great success in many areas of physics. In particular, it is known that machine learning approaches are well suited to classification problems. Classifiers are able to sort input data based on potentially subtle or hard to define features, which may be obscured by noise. Famously, classifiers may be trained to recognise faces, based on many training examples, but without needing to be told the features of a face. In physics, machine learning has been successfully used in classification of galaxy morphology \cite{1807.00807} and jets in particle colliders \cite{1904.08986}, to name but two. Machine learning techniques have also been proposed for anomaly detection in X-ray spectra \cite{2019MNRAS.487.2874I}.

We will focus in this work on a supervised learning approach, in which we train our classifiers with labelled sample data -- in this case spectra simulated with and without the effects of ALPs. We note in passing that  \emph{unsupervised} learning, in which the classifiers are not given training data, may also have potential for physics discovery \cite{1807.06038,1812.02183,1904.04200}.

\section{The Problem}
\label{sec:problem}
We search for ALP induced modulations in the spectra of point sources in or behind galaxy clusters, as observed by the {\it Chandra} X-ray telescope \cite{1005.4665}. We process each observation using CIAO 4.8.1 \cite{ciao}, stacking observations from the same source and subtracting the cluster background. We consider the energy range $1 - 5$ keV\footnote{The only exception being NGC1275 where we consider $0.8 - 5$ keV.}, where {\it Chandra} has a high effective area. Our spectra are significantly impacted by {\it Chandra's} energy resolution of $150$ eV (FWHM). In effect, we observe the true spectrum convolved with a Gaussian of FWHM $150$ eV. This will partially blurr any ALP induced features. Furthermore, our spectra will suffer Poisson noise, with amplitude determined by the observation time. ALP-induced oscillations could potentially hide within this Poisson noise. We model both these effects directly by simulating fake data using the $\it Sherpa$ software \cite{astro-ph/0108426}, as described below. 

\begin{figure}
\centering \includegraphics[width=0.6\textwidth]{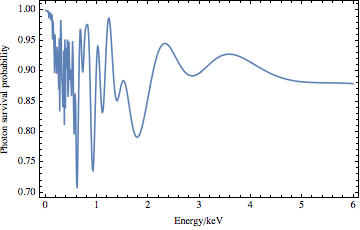}
\caption{The photon survival probability induced by the presence of ALPs with $g_{a \gamma \gamma} = 10^{-12} \, {\rm GeV}^{-1}$ and a realisation of the magnetic field of A1367.}
\label{survival}
\end{figure}

We seek to distinguish between two models for our observed flux - $F(E) = A(E)$ and $F(E) = A(E) P_{\gamma \to \gamma}(E,g_{a \gamma \gamma},{\bf B})$. $A(E)$ is the point source's spectrum assuming standard astrophysics with no ALPs, described in more detail below. $P_{\gamma \to \gamma}(E,g_{a \gamma \gamma},{\bf B})$ is the photon survival probability induced by the presence of ALPs with a coupling $g_{a \gamma \gamma}$ to the photon and the magnetic field $\bf B$ along the line of sight to the source. An example of such a photon survival probability is shown in Figure~\ref{survival}.  Note that oscillations in a spectrum could also result from a different mechanism, such as mis-modelling atomic lines, or instrumental effects such as pileup. \\

\begin{figure}
\hspace*{-0.5in}
\includegraphics[width=.6 \textwidth]{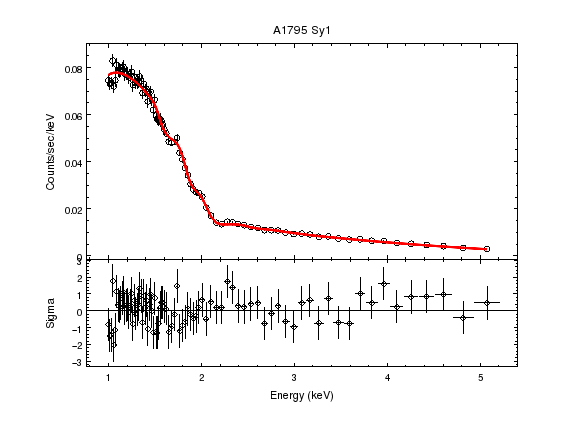}
\includegraphics[width=.6 \textwidth]{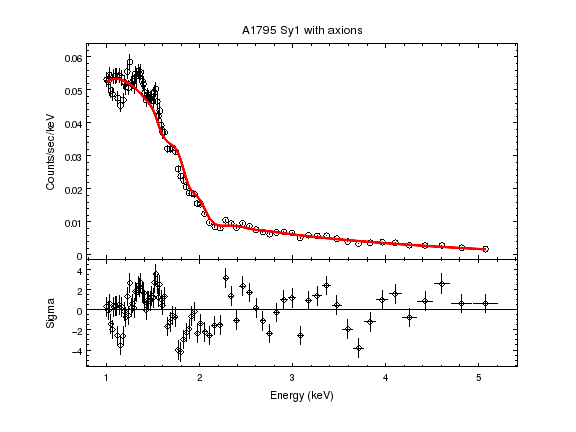}
\caption{Left: the observed spectrum of the Seyfert galaxy 2E3140 in the galaxy cluster A1795 fitted with an absorbed power law. Right: the same spectrum multiplied by the photon survival probability for a realisation of the A1795 magnetic field and assuming the existence of ALPs with $g_{a \gamma \gamma} = 5 \times 10^{-12} \, {\rm GeV}^{-1}$.} 
\label{spectra}
\end{figure}

Figure~\ref{spectra} shows the observed spectrum of the Seyfert galaxy 2E3140 in the galaxy cluster A1795, and its simulated spectrum assuming the existence of ALPs with $g_{a \gamma \gamma} = 5 \times 10^{-12} \, {\rm GeV}^{-1}$ and a particular realisation of the A1795 magnetic field. We see that ALPs induce characteristic oscillations in the residuals, with larger wavelengths than those from Poisson fluctuations alone. The magnitude and power spectrum of {\bf B} for a particular galaxy cluster is inferred from observations of Faraday rotation measures and synchrotron emission. However, the specific configuration of {\bf B} along the line of sight to the point source is unknown, and represents a large set of nuisance parameters in our attempts to constrain $g_{a \gamma \gamma}$. It is important to realise that the form of $P_{\gamma \to \gamma}(E,g_{a \gamma \gamma},{\bf B})$ depends heavily on the precise form of ${\bf B}$. For example, for a different magnetic field configuration, the peaks and troughs of $P_{\gamma \to \gamma}(E,g_{a \gamma \gamma},{\bf B})$ would occur at different energies. However, some features of the spectrum, such as the increasing wavelength of the oscillations with increasing energy, are generic. This fact makes it in principle possible to distinguish characteristically {\it ALP-like} features. This has been explored by considering the spectra in Fourier space in~\cite{1808.05916}. This work also considers using ML to place bounds on ALPs. Here we extend this effort, and find that ML can accelerate the search for ALPs.

Previous work has also searched for ALP induced oscillations in point source spectra, placing bounds on ALPs relying solely on the fact that these oscillations would make the spectra a bad fit to the astrophysics only model. Such search strategies have already placed leading bounds on low mass ALPs. A recent analysis of {\it Chandra} High-Energy Transmission Grating observations, which offer a higher energy resolution, achieves $g_{a \gamma \gamma} \lesssim 6 - 8 \times 10^{-13} \, {\rm GeV}$ \cite{1907.05475}. Several studies of data taken without the grating yield $g_{a \gamma \gamma} \lesssim 1.5 \times 10^{-12} \, {\rm GeV}$ \cite{1605.01043,1703.07354,1704.05256,Chen:2017mjf}.

However, searches based on a $\chi^2$ test or similar do not take into account the distinctive characteristics of ALP induced oscillations. We can therefore hope that our ALP searches will be improved by using machine learning to seek out ALP-like features in the spectra of point sources shining through galaxy clusters.

\section{Astrophysical Systems}
In this work, we will use {\it Chandra} observations of a number of point sources located in or behind galaxy clusters as a test bed for the potential of machine learning in searching for ALPs. Our observations were all taken without the High Energy Transmission Grating. We use the point sources considered in \cite{1605.01043,1704.05256}. These are:

\begin{itemize}
\item The AGN NGC1275 at the centre of Perseus.
\item The quasars B1256+281 and SDSS J130001.48+275120.6 shining through Coma.
\item The AGN NGC3862 in A1367.
\item The AGN IC4374 at the centre of A3581.
\item The bright Sy1 galaxy 2E3140 within A1795.
\item The quasar CXOU J134905.8+263752 behind A1795.
\item The central AGN UGC9799 of the cluster A2052.
\end{itemize}

These sources were chosen based on their brightness and observation time with {\it Chandra}. To simulate the ALP-photon interconversion probability for these sources, we require estimates for the magnetic fields in their host clusters. We use the electron density and magnetic field estimates from \cite{1605.01043,1704.05256}. These are taken from published estimates derived from thermal emission and Faraday rotation measures respectively \cite{Churazov:2003hr,Taylor:2006ta,1002.0594,1201.4119,1703.08688,0912.3930,astro-ph/9712293,ref38,astro-ph/0005224,ref40,astro-ph/0211027,1412.4558,astro-ph/0410154,1003.2719} and from extrapolation from similar clusters when no such estimates are available. We only find competitive bounds from the AGN in A1367, the quasar behind A1795 and the Seyfert 1 galaxy in A1795. These are the same sources for which competitive bounds are obtained using conventional statistical methods in \cite{1704.05256}. The constraining ability of different sources is driven primarily by the number of photon counts available for each source. In the case of NGC1275, a large number of photon counts are available, but the spectrum displays highly significantly anomalies \cite{1605.01043} which limits its constraining power. These are probably a result of instrumental effects. For the rest of the main body of the paper, we will restrict our discussion to the three sources for which competitive bounds are found. We will return to discuss the potential of NGC1275 in appendix~\ref{app:NGC1275}.

\section{Datasets}
\label{sec:datasets}
To train our classifiers, we require many simulated data sets, based on the measured spectrum of each point source, simulated both with and without the effects of ALPs. In the former case, the simulated data sets will differ in the realisation of the magnetic field along the line of sight to the source, and in the realisation of the Poisson errors in each bin. In the latter case, only the Poisson errors will differ between data sets. To this end, we generated  $1000$ magnetic field realisations for each ALP coupling from $0.1-2.0\times 10^{-12}$ GeV in steps of $0.1\times 10^{-12}$ GeV. Crucially, we generate a different set of magnetic fields for each $g_{a \gamma \gamma}$. Each magnetic field realisation is composed of $\mathcal{O}(100)$ cells (chosen to match the physical size of the cluster), with cell sizes drawn from a power law distribution. Within each cell, the magnetic field is constant with a randomly chosen direction, and an amplitude set by its distance from the cluster centre. The distribution of cell sizes and the radial fall off is different for each cluster, as described in \cite{1704.05256}. For each such magnetic field, we simulate the photon survival probability as a function of energy assuming the presence of ALPs with coupling $g_{a \gamma \gamma}$ to the photon, by numerical solution of equation \ref{conversion}.\\

Using Sherpa, we fit the data in the range $1 - 5$ keV from each point source to an power law model with absorption from neutral hydrogen, also allowing a soft thermal component where this improves the fit. This gives us the astrophysics only model $A(E)$ for that source. From these best fit models, we generate fake data sets using the {\it Sherpa} fake data function, in the cases of no ALPs (source spectrum $A(E)$) and for each of the photon survival probabilities simulated (source spectrum $A(E) P_{\gamma \to \gamma}(E,g_{a \gamma \gamma},{\bf B})$). For each case, we generate $10^4$ different fake data sets, which differ from each other in the realisation of Poisson errors in each bin (and underlying magnetic fields). The level of this noise is set by the simulated exposure time, which we set to be the same as that of the actual observation. Generating our fake data in this way takes into account the instrumental response of the telescope, in particular including its energy resolution.

We now seek to compare the spectra simulated with and without the presence of ALPs. In particular, we seek to build a classifier that can distinguish between the ALP and no ALP cases. There are three main differences between the cases with and without ALPs:

\begin{itemize}
\item The spectra with ALPs have overall lower flux.
\item The spectra with ALPs have a higher decrease in flux as we increase energy.
\item The spectra with ALPs display oscillations about the power law model, greater than would be expected from Poisson fluctuations alone, and with increasing wavelength for higher energies. 
\end{itemize}

We may only use the last of these differences in trying to distinguish ALP and non-ALP spectra, as we do not know the intrinsic amplitude or power law index of the source. We therefore cannot train classifiers with our raw simulated spectra and then hope to use them meaningfully on real data. We consider three approaches to this problem. Firstly, we refit each spectrum (both ALPs and no ALPs) to an absorbed power law model, and train our classifiers on the residuals of this fit. Secondly, we rescale our simulated data to remove the first two effects. We do this using the following `upscaling' procedure:

\begin{enumerate}

\item For each coupling $g_{a \gamma \gamma}$, we calculate the average photon survival probability per bin, averaging over each magnetic field realisation. This gives us a function $P^{\rm av}_{\gamma \to \gamma}(E, g_{a \gamma \gamma})$, evaluated at each energy bin.

\item We find the inverse of the average photon survival probability $P^{\rm av^{-1}}_{\gamma \to \gamma}(E, g_{a \gamma \gamma})$. In practice, this is found by simply taking the inverse of the value of $P^{\rm av}_{\gamma \to \gamma}(E, g_{a \gamma \gamma})$ in each energy bin.

\item We generate fake data with ALPs using the source model \\
$A(E) P_{\gamma \to \gamma}(E, g_{a \gamma \gamma}, {\bf B}) P^{\rm av^{-1}}_{\gamma \to \gamma}(E, g_{a \gamma \gamma})$, rather than simply $A(E) P_{\gamma \to \gamma}(E, g_{a \gamma \gamma}, {\bf B})$.

\end{enumerate}

In this way, the average, large scale effects of ALP-photon conversion (the overall decrease, and the increased suppression at high energies) are removed, but the local features that cannot be modeled by standard astrophysics are retained. Thirdly, we use both techniques simultaneously, first performing the upscaling procedure described above and then refitting to an absorbed power law and training our classifiers with the residuals.

\section{Classifiers}

We report results on the following ML classifiers \cite{hastie01statisticallearning, scikit-learn} on labelled spectra simulated with and without the presence of ALPs, as described above:

\begin{itemize}
\item Gaussian Naive Bayes (GaussianNB)
\item Quadratic Discriminant Analysis (QDA)
\item Random Forest Classifier (RFC)
\item Ada Boost Classifier (ABC)
\item Gaussian Process Classifier (GPC)
\item Decision Tree Classifier (DTC)
\item K Neighbours Classifier (KNC)
\item Support Vector Machine (SVM)
\end{itemize}

Each classifier $\mathcal{C}_{g}$ is trained to distinguish between spectra with no ALPs and those with ALPs with coupling $g$. \footnote{One could also build multi-category classifiers trained with a range of couplings. In this proof-of-prinicple paper, we restrict ourselves to the simpler two-category classifiers. These are generally more straight forward to train and assess. We note also that the methods presented here are only appropriate to setting bounds on the ALP parameters, and not for making a discovery. One could certainly use these classifiers as a discovery tool, but the look elsewhere effect would need to be carefully taken into account.} We split our dataset of simulated data samples into training and test sets. We have performed numerical experiments using varying sizes of training sets $N=(3600, 4500,6400,8000)$ for the classifiers. We check that the results do not vary significantly with $N$. For the GPC, KNC and SVM, we did not obtain competitive bounds. We present our results for the other five classifiers below.

We can understand the inefficacy of the GPC, KNC and SVM classifiers as follows. The Gaussian Process Classifier assumes that the classifier function can be modelled as a Gaussian Process - as assumption that is not well justified this case. The perforamance of the Support Vector Machine is highly dependent on the choice of kernel. As our other classifiers give good performance already, we have not optimized the SVM kernel. The K-Neighbours Classifier is a relatively simple classifier that does not generally track higher order features of the data. Such higher order features are essential for characterising ALP-photon oscillations, so it is unsurprising that no bounds are obtained. In general, the performance of a classifier is dependent on the choice of hyper-parameters. Such a choice of hyper-parameters corresponds to a different choice of bias on the model space. This biasing provides a reason for the observed difference in performance in the classifiers. To see the general viability of using classifiers, we would like to stress that we observe reasonable performance across various classifiers and, as previously explained,  the expected failure of several classifiers.

\section{Classifier Performance}

We have a set of classifiers $\mathcal{C}_g$ trained to classify spectra as containing ALP-induced oscillations or no ALP-induced oscillations. We have trained these classifiers either on residuals when data is fit with a power law, or by using the `upscaled' ALP data as described above, or using the residuals from upscaled data. For classifer $\mathcal{C}_g$, the training data with ALPs was generated assuming an ALP-photon coupling $g$, and using a range of randomly generated magnetic fields. We can test the performance of these classifers using separate sets of test data, generated without ALPs or with ALPs at a range of $g$ values. For example, all classifiers should show the same behaviour for data simulated with very low values of $g$ as for data simulated without ALPs.

\begin{figure}
\hspace{-1cm}
\includegraphics[width=0.55\textwidth]{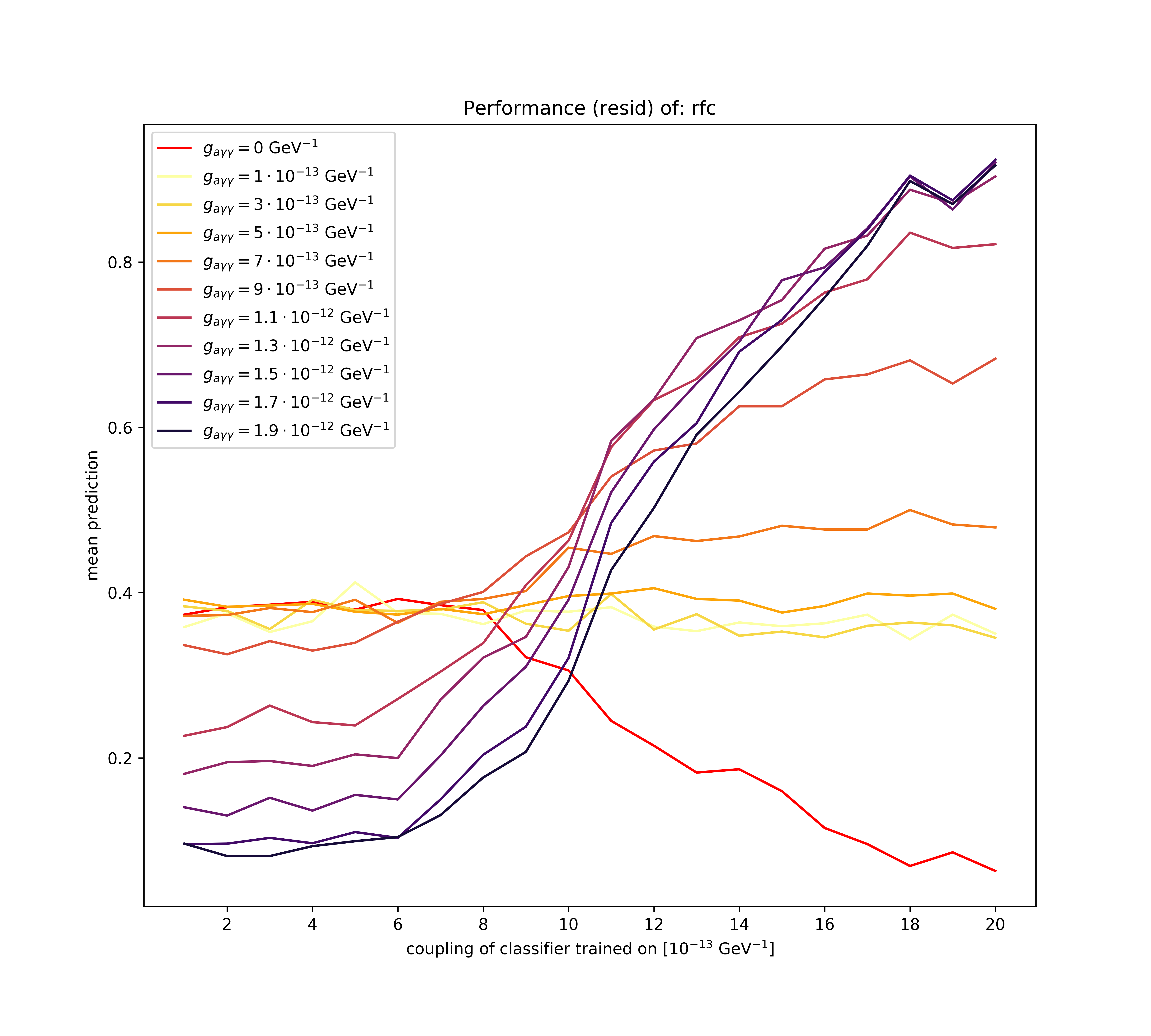}
\includegraphics[width=0.55\textwidth]{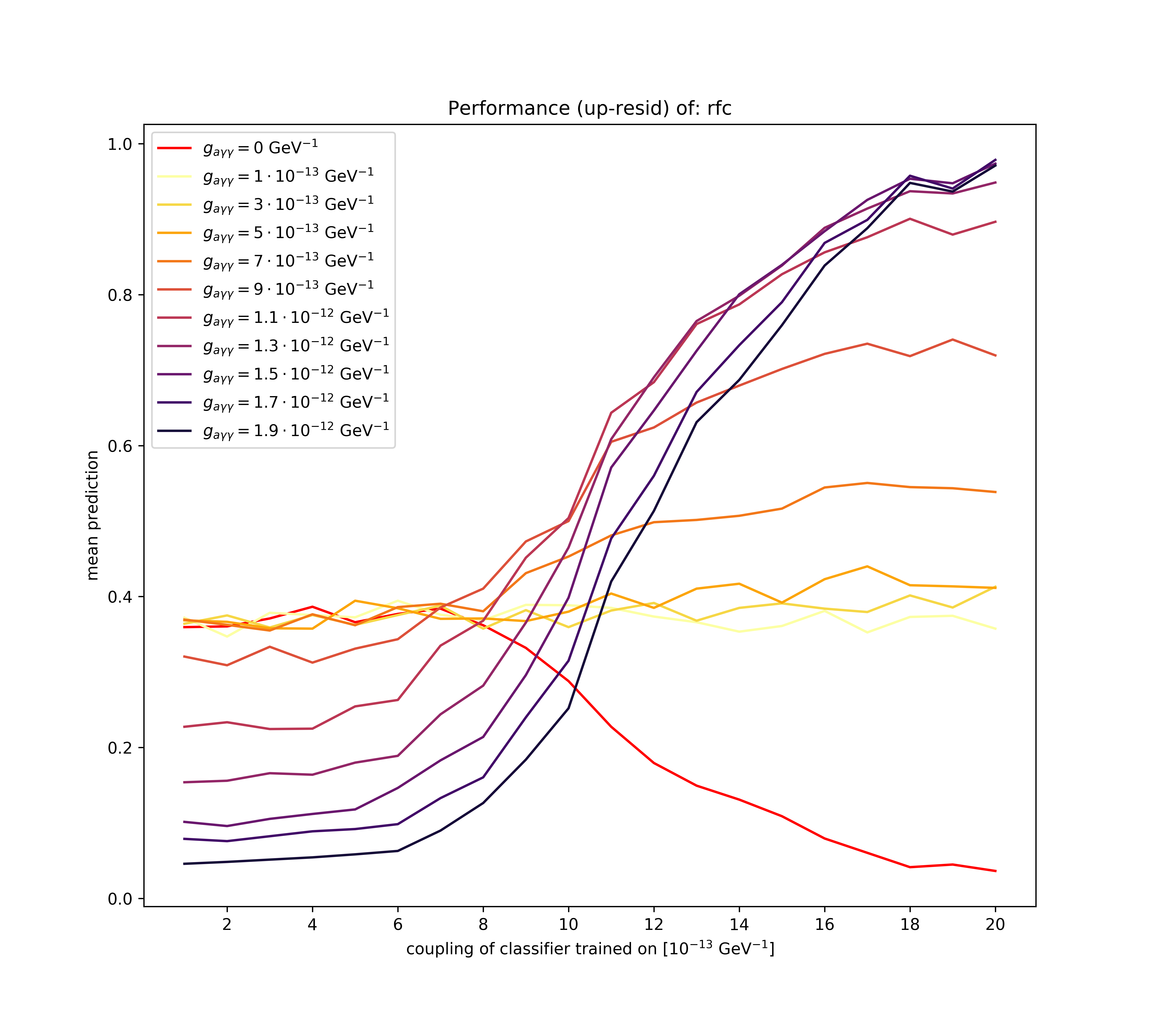}
\caption{The performance of the RFC classifiers trained on residuals (left) and up-scaled residuals (right). The classifier $C_g$ is trained to distinguish data with no ALPs from data with ALPs with coupling $g$, where $g$ is shown on the x axis. We query each classifier with data generated with the full range of coupling values $g_{\rm query}$. Each coloured line corresponds to a differet $g_{\rm query}$. The legend shows $g_{\rm query}$, with $g_{\rm query} = 0$ corresponding to no ALPs. The y axis shows the mean output for a classifier $C_g$ queried by data with coupling $g_{\rm query}$, where $0$ corresponds to no ALPs and $1$ to ALPs.}
\label{GaussianNBResid}
\end{figure}

\begin{figure}
\hspace{-1cm}
\includegraphics[width=0.55\textwidth]{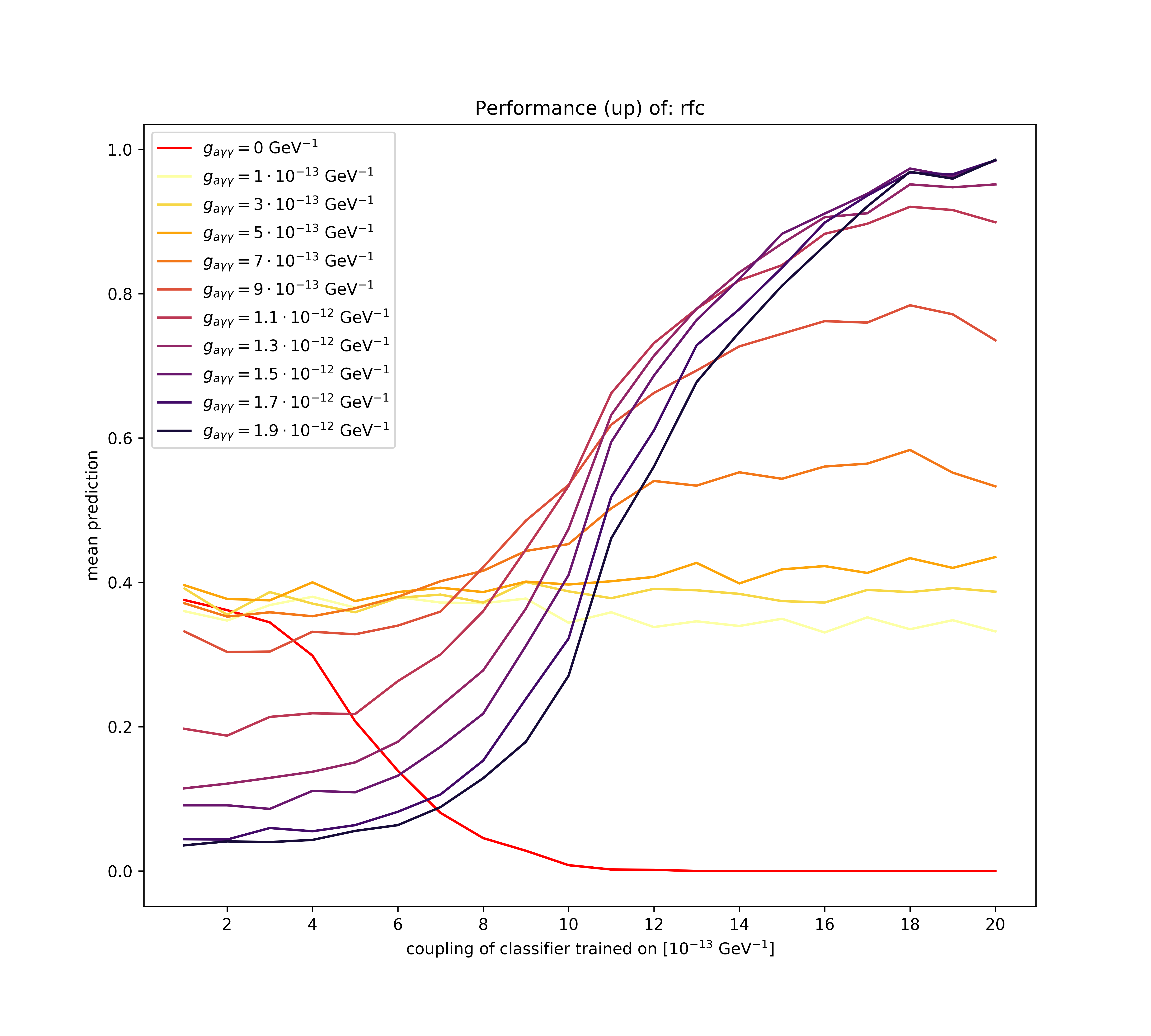}
\raisebox{0.2\height}{\includegraphics[width=0.55\textwidth]{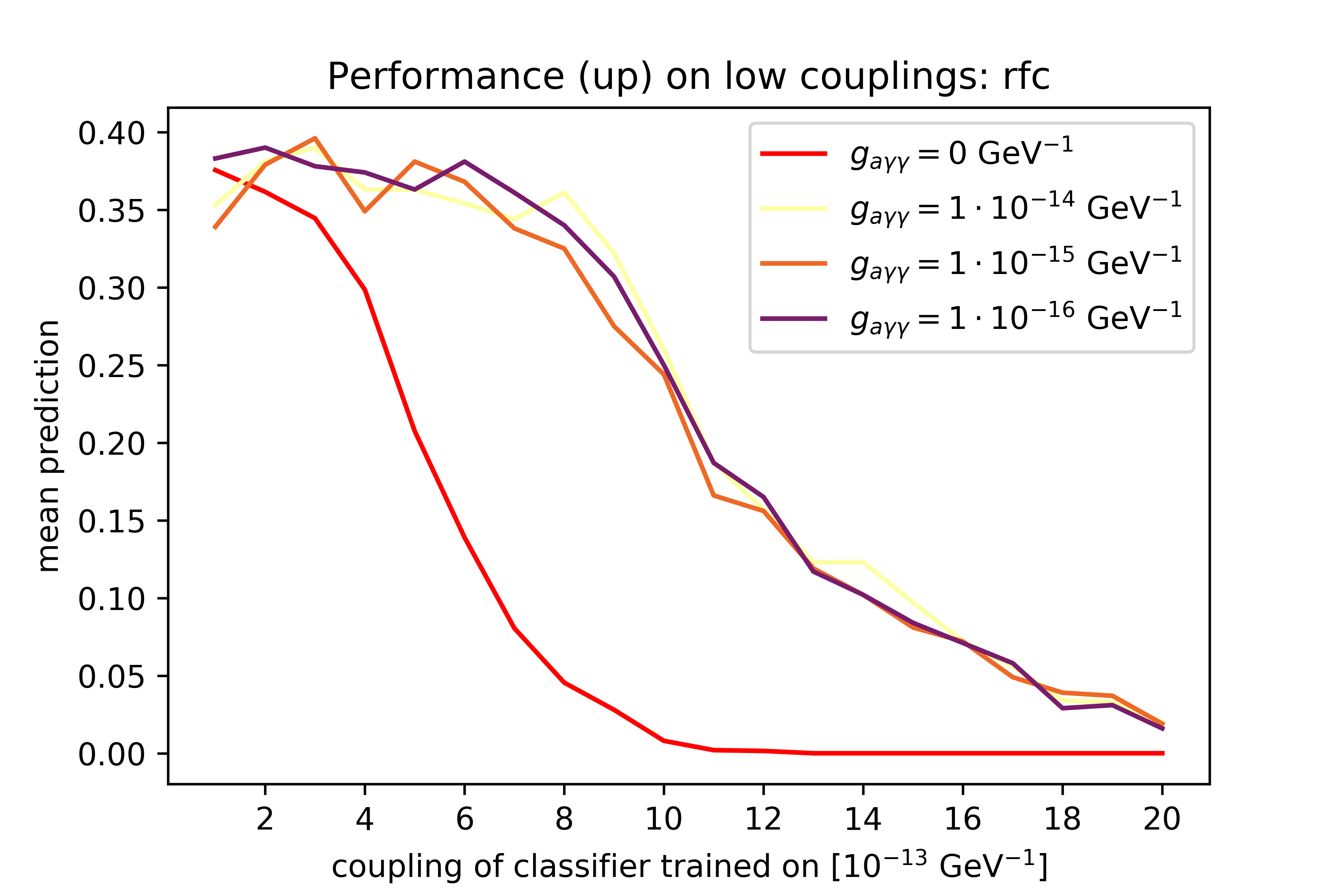}}
\caption{The performance of the RFC classifiers trained on upscaled data (for couplings as in Figure~\ref{GaussianNBResid} (left) and for very low values of $g_{\rm query}$ (right)). The classifier $C_g$ is trained to distinguish data with no ALPs from data with ALPs with coupling $g$, where $g$ is shown on the x axis. We query each classifier with data generated with the full range of coupling values $g_{\rm query}$. Each coloured line corresponds to a different $g_{\rm query}$. The legend shows $g_{\rm query}$, with $g_{\rm query} = 0$ corresponding to no ALPs. The y axis shows the mean output for a classifier $C_g$ queried by data with coupling $g_{\rm query}$, where $0$ corresponds to no ALPs and $1$ to ALPs.}
\label{GaussianNBUp}
\end{figure}

Figures \ref{GaussianNBResid} and \ref{GaussianNBUp} show the performance of our classifiers $\mathcal{C}_g$ when queried with data simulated with different ALP-photon couplings $g_{\rm query}$ for the Sy1 galaxy 2E3140 within A1795. We see that for classifiers trained with sufficiently high $g$ values, data simulated with ALPs is mostly classified as such, while data simulated without ALPs is also mostly classified correctly. For all three data processing methods, we see a clear separation between ALP and no ALP data. The performance of the other classification algorithms follows a similar pattern. Interestingly, we also find that when classifiers trained with a relatively low value of $g$ ($\mathcal{O} (10^{-13}) \, {\rm GeV}^{-1}$) are queried with data simulated with a high value of $g_{\rm query}$  ($\mathcal{O} (10^{-12}) \, {\rm GeV}^{-1}$), the result is usually `No ALPS'. This suggests that there are significant qualitative differences between the high and low $g$ regimes that the classifiers are picking up on. This behaviour does not affect our test statistic, defined below, and therefore does not affect our bounds. 

\begin{figure}
\hspace{-1cm}
\includegraphics[width=0.55\textwidth]{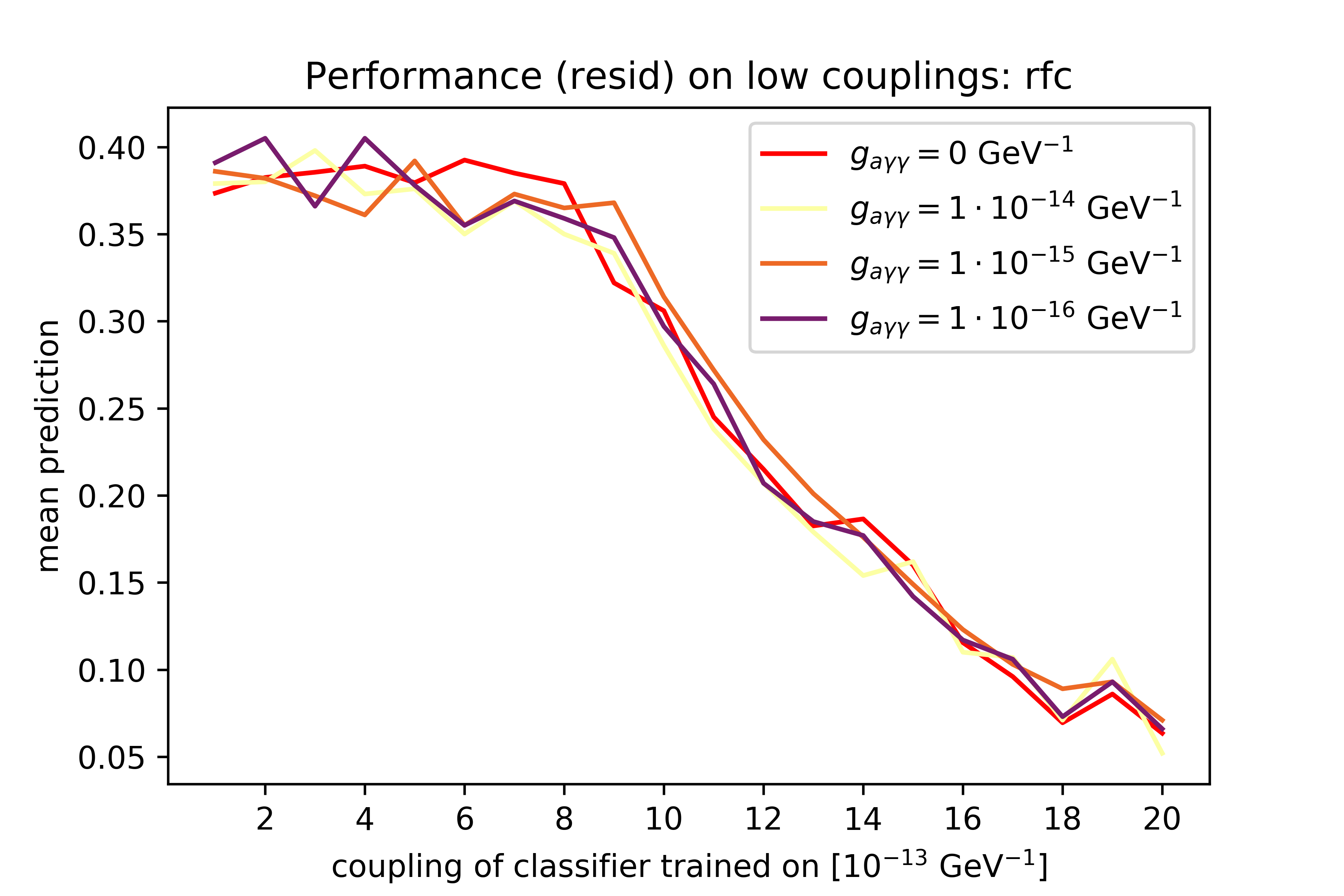}
\includegraphics[width=0.55\textwidth]{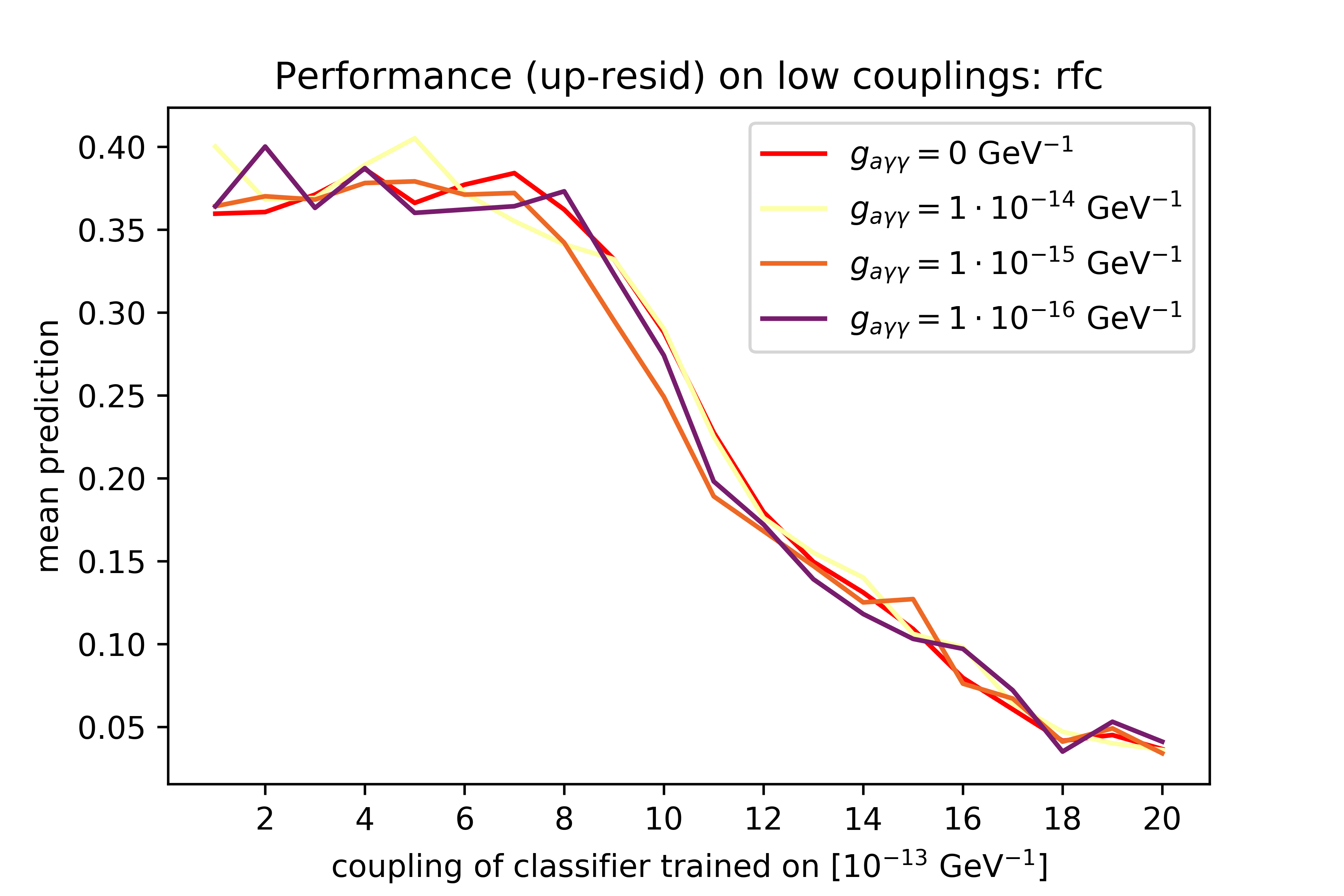}
\caption{As figure \ref{GaussianNBResid} but for very low values of $g_{\rm query}$.}
\label{GaussianNBResidLow}
\end{figure}

Figures \ref{GaussianNBUp} and \ref{GaussianNBResidLow} (right) show the performance of our RFC classifiers for very low values of $g_{\rm query}$. We expect this to be the same as their performance for  $g_{\rm query} = 0.$ We see that this is true for the classifiers trained with residuals and upscaled residuals, but not for the classifiers trained with upscaled data. This is also the case for the other classification algorithms. The cause of this bias in the upscaled classifiers is not known. We therefore do not use the upscaled classifiers for setting bounds.\footnote{The bounds obtained from the upscaled classifiers are significantly better than those from the residual classifiers.}

We can also use our classifiers on real data, and hence obtain a bound on $g_{a \gamma \gamma}$. We input our real data to each of our classifiers $\mathcal{C}_g$. The output of each $\mathcal{C}_g$ will be either \emph{ALPs} or \emph{No ALPs}. For very high values of $g$, assuming the data does not contain such ALPs, we expect the classifier to return \emph{No ALPs} a very high proportion of the time. At intermediate values of $g$, on the boundary of what would be detectable, we might expect the classifier to return \emph{No ALPs} the majority of the time. Furthermore, if ALPs actually are present in the data with coupling $g = g_{a \gamma \gamma}$, we would expect classifiers trained with couplings close to $g_{a \gamma \gamma}$ to return \emph{ALPs} most of the time. This is the effect we want to use to place bounds on $g_{a \gamma \gamma}$.\\

\section{Bounds}

Having established that our classifiers can distinguish between observations simulated with and without ALPs, we now seek to use them to place bounds on the ALP-photon coupling.\\

\noindent We define a test statistic for a data set $\mathcal{D}$: 
\begin{center}
${\rm TS}_{\mathcal{D}} = $  highest value of $g$ such that $\mathcal{C}_g$ classifies $\mathcal{D}$ as \emph{ALPs}
\end{center}
For example, more noisy data, in which it is easier for ALPs to hide, and hence harder to distinguish the ALP and no ALP cases, will have a higher ${\rm TS}_{\mathcal{D}}$. To place bounds on ALPs we consider the null hypothesis:
\begin{center}
$H_0$: ALPs exist with $g = g_{\rm null}$.
\end{center}

We now find the null distribution of ${\rm TS}_{\mathcal{D}}$ by Monte Carlo. We generate $2000$ fake data sets (i.e. spectra) $\{ \mathcal{D}^i (g_{\rm null}) \} $ assuming ALPs with $g = g_{\rm null}$ with different magnetic field realisations. We find ${\rm TS}_{\mathcal{D}^i(g_{\rm null})}$ for each fake data set in $\{ \mathcal {D}^i(g_{\rm null}) \}$. If 95 \% of the ${\rm TS}_{\mathcal{D}^i(g_{\rm null})}$ are higher than the test statistic for the real data, $g_{\rm null}$ is excluded at the 95 \% confidence level.\\

For each data set, we check if the null distribution of ${\rm TS}_{\mathcal{D}}$ has an approximately Gaussian form. There are two circumstances in which this might not happen:

\begin{itemize}
\item If the training data is so noisy that no value of $g$ (or no tested value of $g$) has a significant effect on the data, then ${\rm TS}_{\mathcal{D}}$ will just take a random value for each fake data set, whatever the value of $g_{\rm null}$. The resulting distribution will clearly not be Gaussian. The training data has the same noise level as the real data we intend to classify. Physically, in this situation the data is too noisy to place any bounds on $g$.

\item We have trained classifiers with values of $g$ so high that the conversion probability has become saturated. When $g_{\rm null}$ is also high enough that the conversion probability is saturated,  ${\rm TS}_{\mathcal{D}}$ will just be the highest value of $g$ we happened to use for training. Physically, this is because it is not possible to distinguish between different values of $g$ that both saturate the conversion probability. In this case, we can still place upper bounds on $g$ using the lower tail of the null distribution.
\end{itemize}

\begin{figure}
\centering \includegraphics[width=0.55\textwidth]{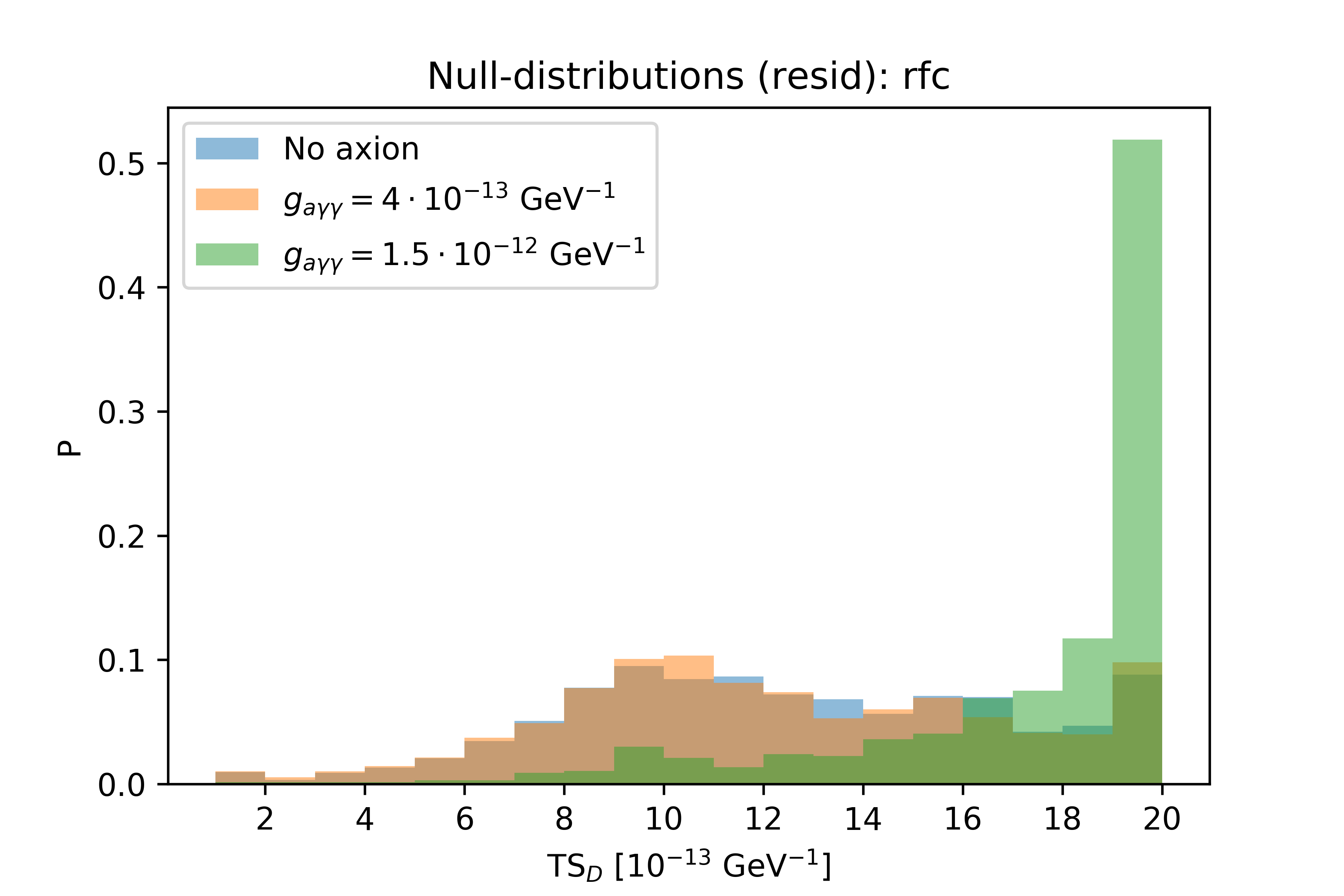}
\caption{Null distribution for the bright Sy1 galaxy 2E3140 within A1795 for no ALPs, low couplings (indistinguishable from no ALPs), and larger couplings of residual types.}
\label{fig:nulldistribution}
\end{figure}

Figure~\ref{fig:nulldistribution} shows a null distribution plot for the bright Sy1 galaxy 2E3140 within A1795 for no ALPs, low couplings (indistinguishable from no ALPs), and larger couplings.

In detail, our bounds procedure for residual classifiers is as follows. The bounds procedure for classifiers using the upscaled residuals is analogous.

\begin{enumerate}
\item Choose a set of $g$ values with which to build classifiers. For example, $g_C = \{1 - 20 \} \times 10^{-13} \, {\rm GeV}^{-1}$.

\item Simulate photon survival probabilities $P^{\rm train}_j(E, g_C)$ for each value of $g_C$ considered and for 800 different magnetic field configurations $\{ B^{\rm train}_j \}$. 

\item Fit the real data with an absorbed power law model, giving a best fit power law $F_{\rm fit}(E)$.

\item For each simulated photon survival probability $P^{\rm train}_j(E, g_C)$, simulate $10$ fake data sets using the exposure and background from the real data and a source spectrum $P^{\rm train}_j(E, g_C) \times F_{\rm fit}(E)$. We therefore have $8000$ fake data sets for each $g_C$.

\item Fit each such fake data set with an absorbed power law, allowing the parameters to vary freely again. Save the residuals from each fit $R^{\rm train}_j(E, g_C)$.

\item Simulate $8000$ fake data sets with no ALPs, i.e. with source spectrum $F_{\rm fit}(E)$. Fit each of these fake data sets with absorbed power law, again allowing the parameters to vary freely.  Save the residuals to from each fit $R^{\rm train}_j(E, 0)$.

\item For each $g_C$, train a classifier $\mathcal{C}_{g_C}$ to distinguish $R^{\rm train}_j(E, g_C)$ from $R^{\rm train}_j(E, 0)$  -- i.e.~to distinguish residuals from data including ALPs with coupling $g_C$ from residuals from data containing no ALPs.

\item Now choose a value of $g$, $g_{\rm null}$, to attempt to exclude. It is not necessary for $g_{\rm null}$ to be equal to any of the $g_C$.

\item Simulate photon survival probabilities $P_j(E, g_{\rm null})$ for $200$ different magnetic field configurations $\{ B_j \}$. These must be different from the magnetic field configurations used for the training data.

\item For each simulated photon survival probability $P_j(E, g_{\rm null})$, simulate $10$ fake data sets using the exposure and background from the real data and a source spectrum $P_j(E, g_{\rm null}) \times F_{\rm fit}(E)$. We therefore have $2000$ fake data sets with $g = g_{\rm null}$.

\item Fit each such fake data set with an absorbed power law, allowing the parameters to vary freely again. Save the residuals from each fit $R_j(E, g_{\rm null})$.

\item Feed each $R_j(E, g_{\rm null})$ to each of the classifiers $C(g_C)$. For each $R_j(E, g_{\rm null})$, record the highest value of $g_C$ for which the corresponding classifier $\mathcal{C}_{g_C}$ returned a verdict of ALPs. We call this value $TS_j(g_{\rm null})$.

\item The bar chart of the $TS_j(g_{\rm null})$ forms the null distribution of the test statistic defined above under the null hypothesis `ALPs with coupling $g_{\rm null}$ exist' (see Figure~\ref{fig:nulldistribution}).

\item Find the residuals $R_{\rm real}(E)$ when the real data is fit with a power law. 

\item Feed $R_{\rm real} (E)$ to each of the classifiers $\mathcal{C}_{g_C}$. Record the highest value of $g_C$ for which the corresponding classifier $\mathcal{C}_{g_C}$ returned a verdict of ALPs. We call this value $TS_{\rm real}$. This is the test statistic for the real data.

\item If $TS_{\rm real}$ lies in one of the tails of the null distribution, we can exclude the null hypothesis with some degree of certainty. Depending on the tail, this could either be because the real data is much more axiony or much less axiony than the fake data with $g = g_{\rm null}$. If 95 \% of the $TS_j(g_{\rm null})$ are higher than $TS_{\rm real}$, $g \geq g_{\rm null}$ is excluded at the 95 \% confidence level.

\end{enumerate}

\section{Results and Discussion}

\begin{figure}
\hspace{-1cm}
\includegraphics[width=0.55\textwidth]{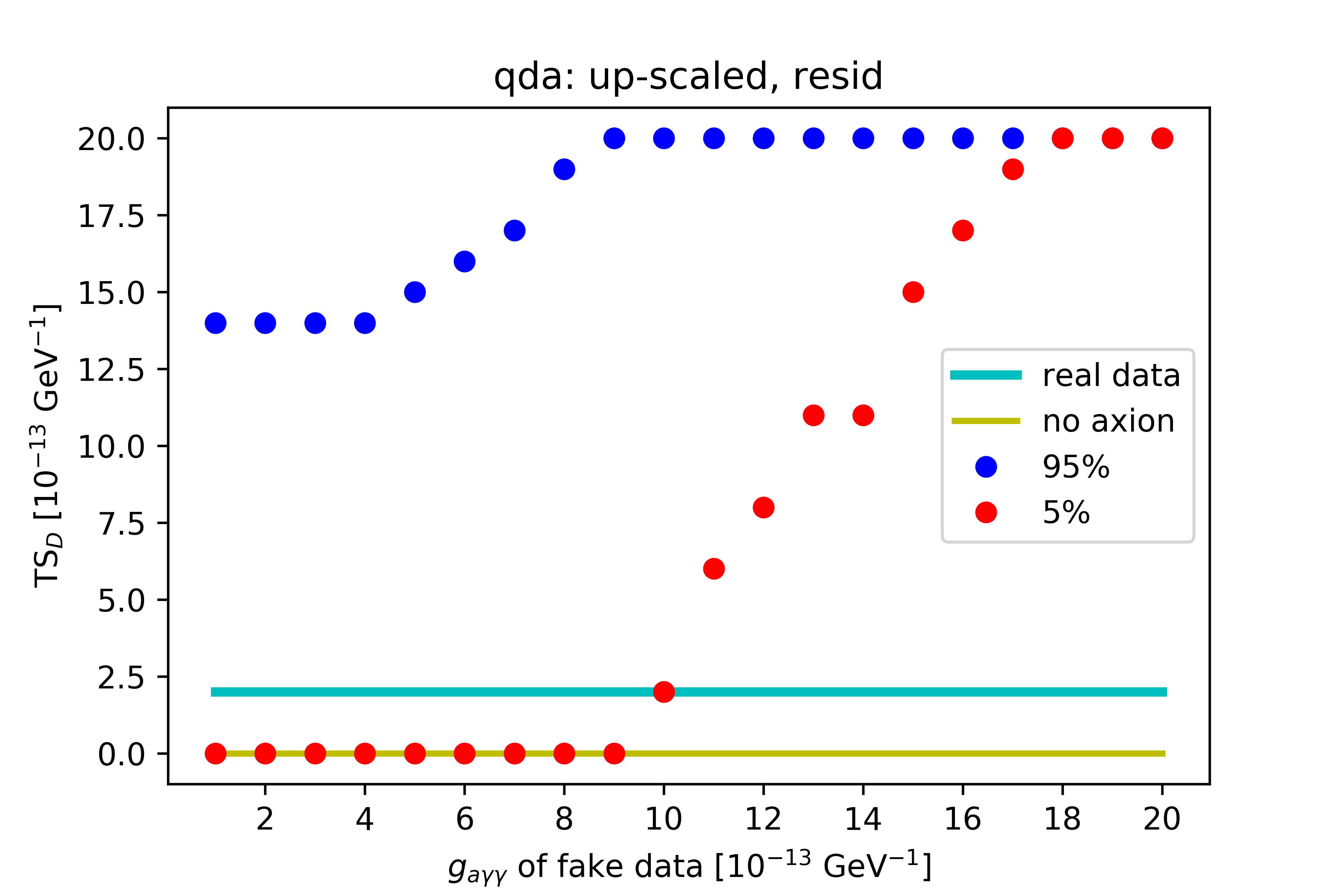}
\includegraphics[width=0.55\textwidth]{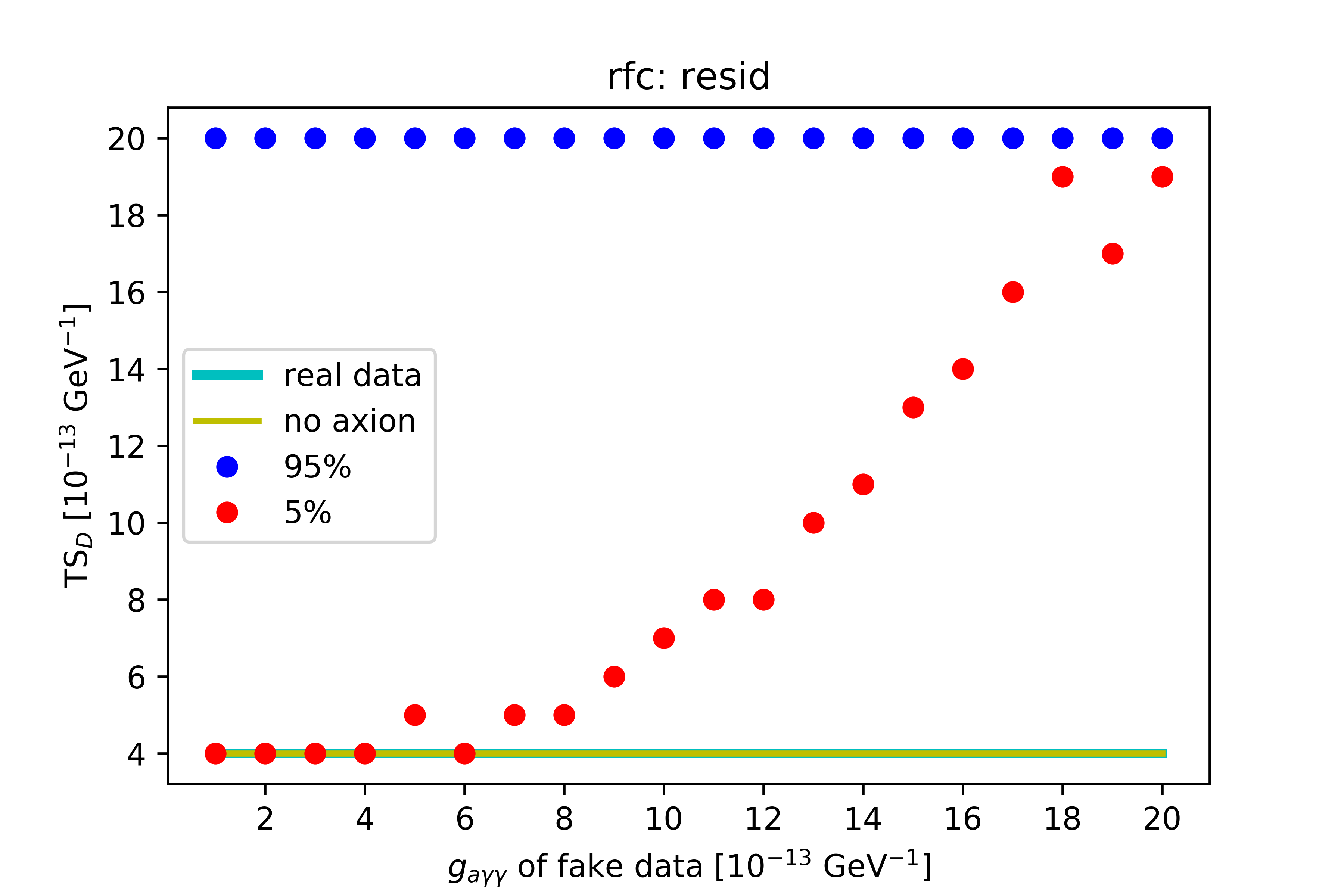}
\caption{Test statistic quartiles for the QDA classifier applied to up-scaled residuals (left) and for the RFC classifier for residuals (right) for the Seyfert 1 galaxy in A1795. We query the classifiers with simulated data with $g_{a \gamma \gamma}$ as shown on the $x$ axis. The $y$ axis shows the $5 \%$ (red)  and $95 \%$ (blue) percentile values of the test statistic defined above. The blue line is the test statistic of the real data and the yellow line is the test statistic of the simulated no-ALP data.}
\label{redDots}
\end{figure}

Figure~\ref{redDots} shows the 5th and 95th percentile values of the test statistic as defined above for the Seyfert 1 galaxy in A1795. We query the classifiers with simulated data with $g_{a \gamma \gamma}$ as shown on the $x$ axis. The $y$ axis shows the $5 \%$ (red)  and $95 \%$ (blue) percentile values of the test statistic. The blue line is the test statistic of the real data. The $95 \%$ confidence limit on $g_{a \gamma \gamma}$ therefore corresponds to the $x$ axis value where the red points cross the blue line. We see that the $5 \%$ percentile test statistic plateaus to a low (in this case zero) value of the test statistic at $g_{a \gamma \gamma} \sim 4.0 \times 10^{-13} \, {\rm GeV}^{-1}$ (RFC, right panel). This corresponds to the maximum constraining power of this source and observation time, in the case that the real spectrum perfectly fits a no ALP model. This is because the classifiers cannot distinguish $g_{a \gamma \gamma} \sim 4.0 \times 10^{-13} \, {\rm GeV}^{-1}$ from $g_{a \gamma \gamma} < 4.0 \times 10^{-13} \, {\rm GeV}^{-1}$ at the $95 \%$ confidence level. 

In this example, the test statistic of the real data is above the $5 \%$ percentile plateau - this is the case for the majority of our sources and classifiers. This shows that the real data appears somewhat `more axiony' than data simulated with very weakly interacting ALPs. This could be due to un-modelled astrophysical or instrumental effects, or simply a result of statistical fluctuations. Therefore, we do not saturate the maximum constraining power of this source with this classifier. In a couple of cases, for the DTC and RFC classifiers with the Sy1 source in A1795, the real data test statistic is lower than the $5 \%$ percentile plateau. This shows that the real data appears somewhat {\it less} `axiony' than data simulated with very weakly interacting ALPs. This is similar to a situation in which the reduced $\chi^2$ value for a data set is less than one. The data is too good a fit to the standard model, for example due to lower than average Poisson fluctuations. In this case, we cannot use our bounds method and so do not report constraints on $g_{a \gamma \gamma}$ for these cases.
\newline

Table \ref{ResultsTable} shows the bounds on $g$ obtained using the method described above. The point source in A1367 and the quasar in A1795 do not consistently give bounds across classifiers. Where these sources fail to give bounds, it is because the test statistic for the real data is rather high -- i.e.~the real data `looks axiony' to the classifier. Given the lack of consistency across classifiers, we do not consider that a reliable bound on $g_{a \gamma \gamma}$ is produced from these sources. On the other hand, applying machine learning classifiers to the Seyfert 1 galaxy within A1795 consistently gives bounds in the range $g_{a \gamma \gamma} \lesssim 0.7 - 1.2 \times 10^{-12} \, {\rm GeV}^{-1}.$ This is in a similar range to the current leading bound \cite{1907.05475} obtained using a significantly higher resolution data set with conventional statistical methods. Furthermore, it improves the previously reported bound in~\cite{1704.05256} of $g_{a \gamma \gamma} \lesssim 1.5 \times 10^{-12} \, {\rm GeV}^{-1}$ for this source.

In this analysis, we have used one-dimensional, domain based magnetic field simulations to obtain the photon survival probabilities. This is standard practice for simulating point sources shining through galaxy clusters in the presence of ALPs \cite{1304.0989,1605.01043,1703.07354,1704.05256,1712.08313,1907.05475}. However, the true magnetic field structure of these clusters is better approximated by a three-dimensional turbulent field simulation, as used in \cite{1603.06978}. We therefore have also tested the response of our classifiers to spectra simulated with a full turbulent field model. \footnote{We thank our anonymous referee for pointing out this issue.} As described in Appendix \ref{app:newmagneticfields}, when using residuals, our classifiers are unable to reliably identify ALP oscillations from the turbulent field simulations. However, when using the upscaled residuals, our classifiers perform well on spectra simulated with the turbulent field model. This discrepancy clearly merits further investigation and will be the subject of future work. For now, we note that the bounds derived here from the fit residuals are very sensitive to the magnetic field structure. These should therefore be taken as a demonstration of the potential of machine learning, rather than as true bounds on $g$. Furthermore, our results suggest that there are important qualitative differences between spectra generated with one-dimensional and three-dimensional magnetic field models. These differences may also prove important for ALP searches in point source spectra using more traditional statistical methods.

In this work we have demonstrated for the first time the use of machine learning techniques for ALP induced oscillations in point source spectra with application to real data. The bounds we obtain are competitive with those obtained using conventional statistical methods. They also represent an improvement on a point source basis, comparing the performance on the same datasets. Machine learning techniques have the potential to increase the reach of ALP searches in both current and future data sets. Future X-ray missions will feature substantially improved energy resolution and effective area, allowing us to probe even smaller values of $g_{a \gamma \gamma}$. The improved energy resolution will reveal the characteristic features of spectral anomalies from ALP-photon interconversion in much greater detail. We therefore anticipate that the gains from machine learning techniques will be larger for future telescopes.


\begin{table}[h]
\begin{tabular}{|c|c|c|c|c|c|c|}
\hline 
 & ABC & DTC & GaussianNB & QDA & RFC \\ 
\hline 
A1367 residuals & 1.9 & none & none & none & none \\ 
\hline 
A1367 upscaled residuals & 2.0 & none & 1.9 & none & none \\ 
\hline 
A1795 Quasar residuals & none & none & 1.7 & none & 1.4 \\ 
\hline 
A1795 Quasar upscaled residuals & none & none & none & none & none \\ 
\hline 
A1795 Sy1 residuals & 1.0 & 0.8 & 1.2  & 1.1 & 0.7 \\ 
\hline 
A1795 Sy1 residuals upscaled & 1.1 & 1.1 & 1.1 & 1.0 & 0.8 \\ 
\hline 
\end{tabular} 
\caption{Bounds on $g$ in units of $10^{-12} \, {\rm GeV}^{-1}$ obtained using machine learning classification algorithms.}
\label{ResultsTable}
\end{table}

\section*{Acknowledgments}
We would like to thank Joe Conlon, Andy Powell, Ben Hoyle and Edward Hughes for valuable discussions.
Part of this research was supported by the Cambridge-LMU partnership programme. This work has been partially supported by STFC consolidated grant ST/P000681/1. FD is supported by a research fellowship from Peterhouse, University of Cambridge.

\appendix

\section{NGC1275}
\label{app:NGC1275}

\begin{figure}[H]
\hspace{-1cm}
\includegraphics[width=0.55\textwidth]{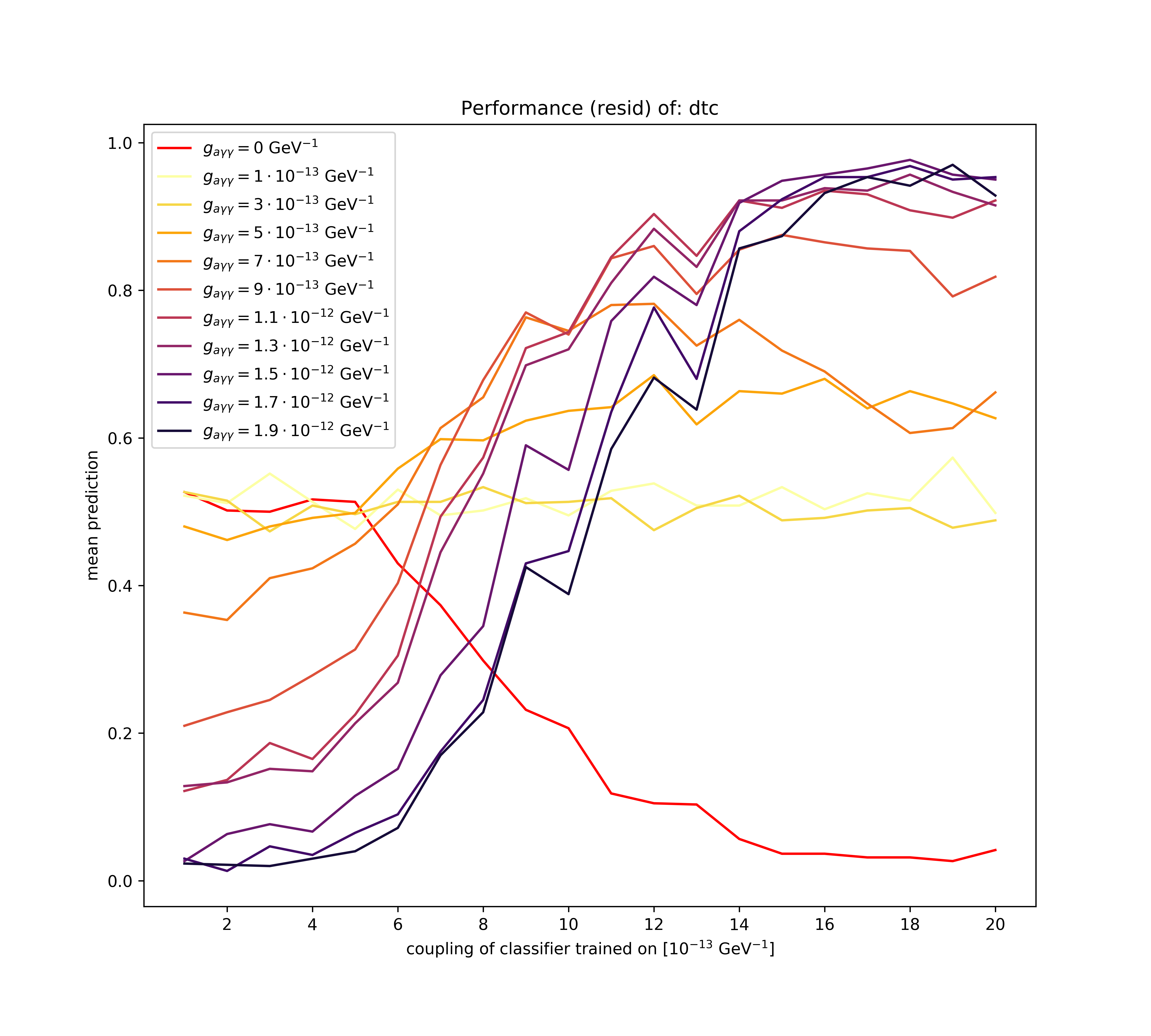}
\includegraphics[width=0.55\textwidth]{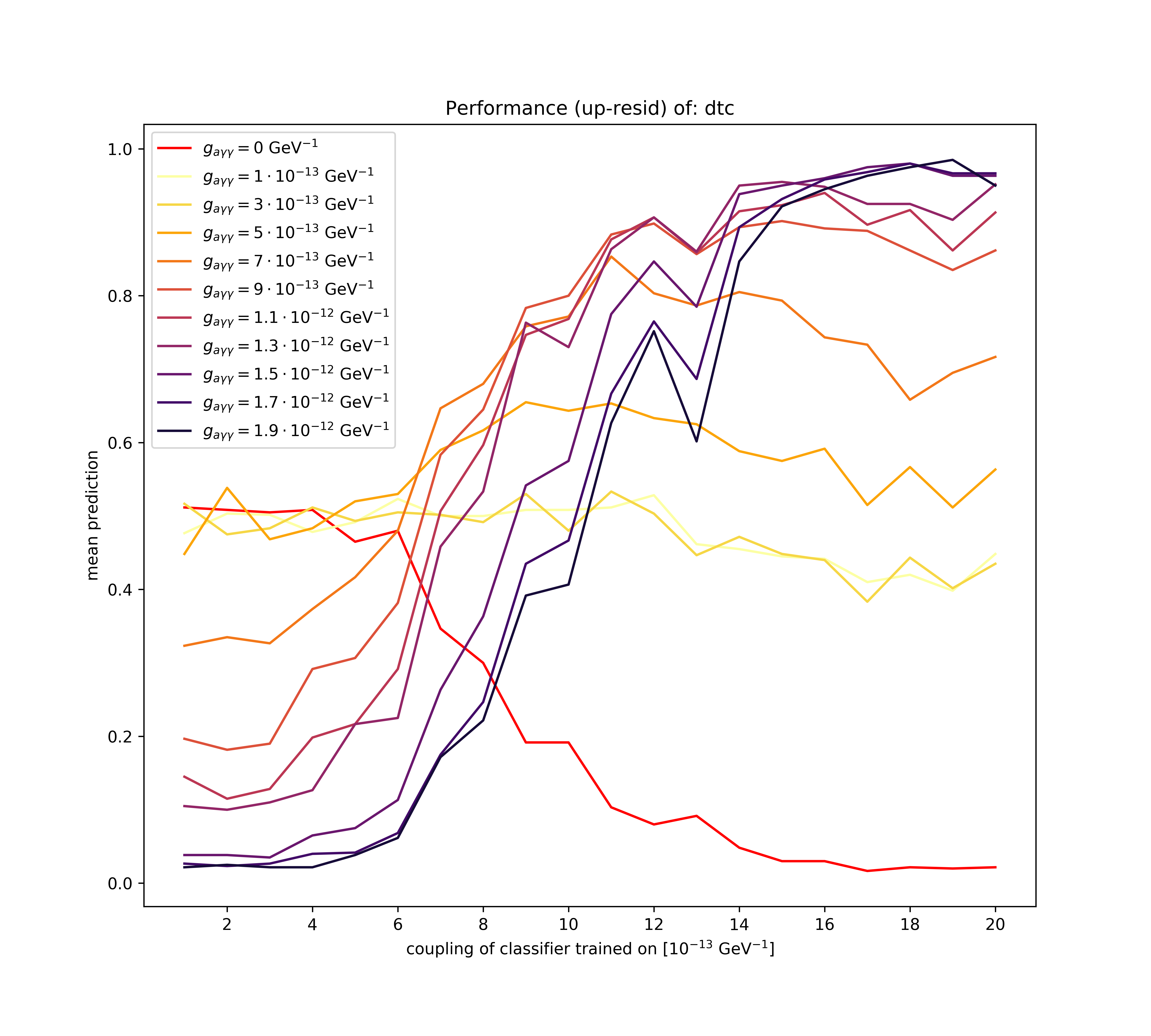}
\caption{The performance of the DTC classifiers trained on residuals (left) and up-scaled residuals (right) for NGC1275, the central AGN of the Perseus galaxy cluster. The classifier $C_g$ is trained to distinguish data with no ALPs from data with ALPs with coupling $g$, where $g$ is shown on the x axis. We query each classifier with data generated with the full range of coupling values $g_{\rm query}$. Each coloured line corresponds to a different $g_{\rm query}$. The legend shows $g_{\rm query}$ in units of GeV, with $g_{\rm query} = 0$ corresponding to no ALPs. The y axis shows the mean output for a classifier $C_g$ queried by data with coupling $g_{\rm query}$, where $0$ corresponds to no ALPs and $1$ to ALPs.}
\label{fig:ngc1275summary}
\end{figure}

\begin{figure}[H]
\hspace{-1cm}
\includegraphics[width=0.55\textwidth]{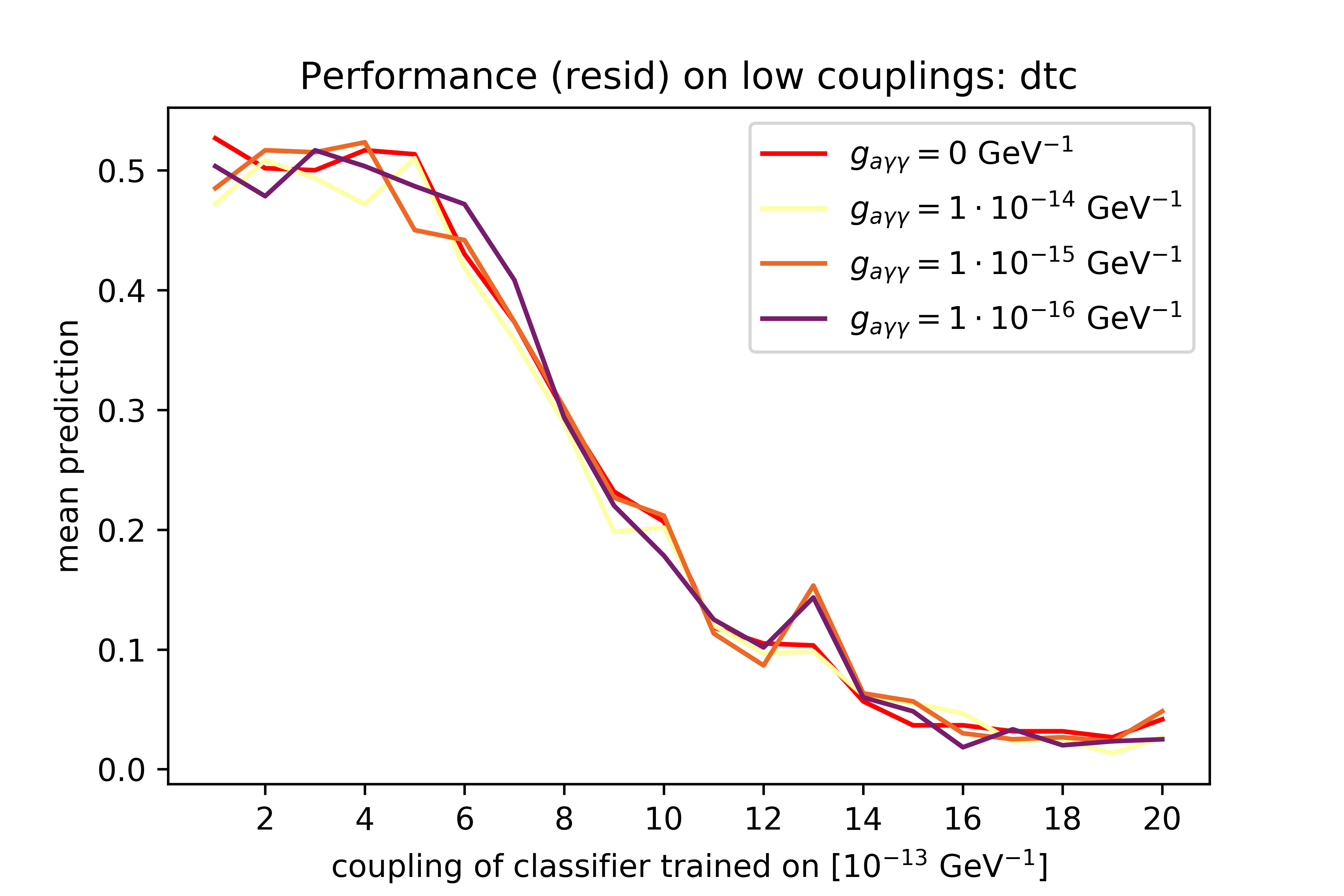}
\includegraphics[width=0.55\textwidth]{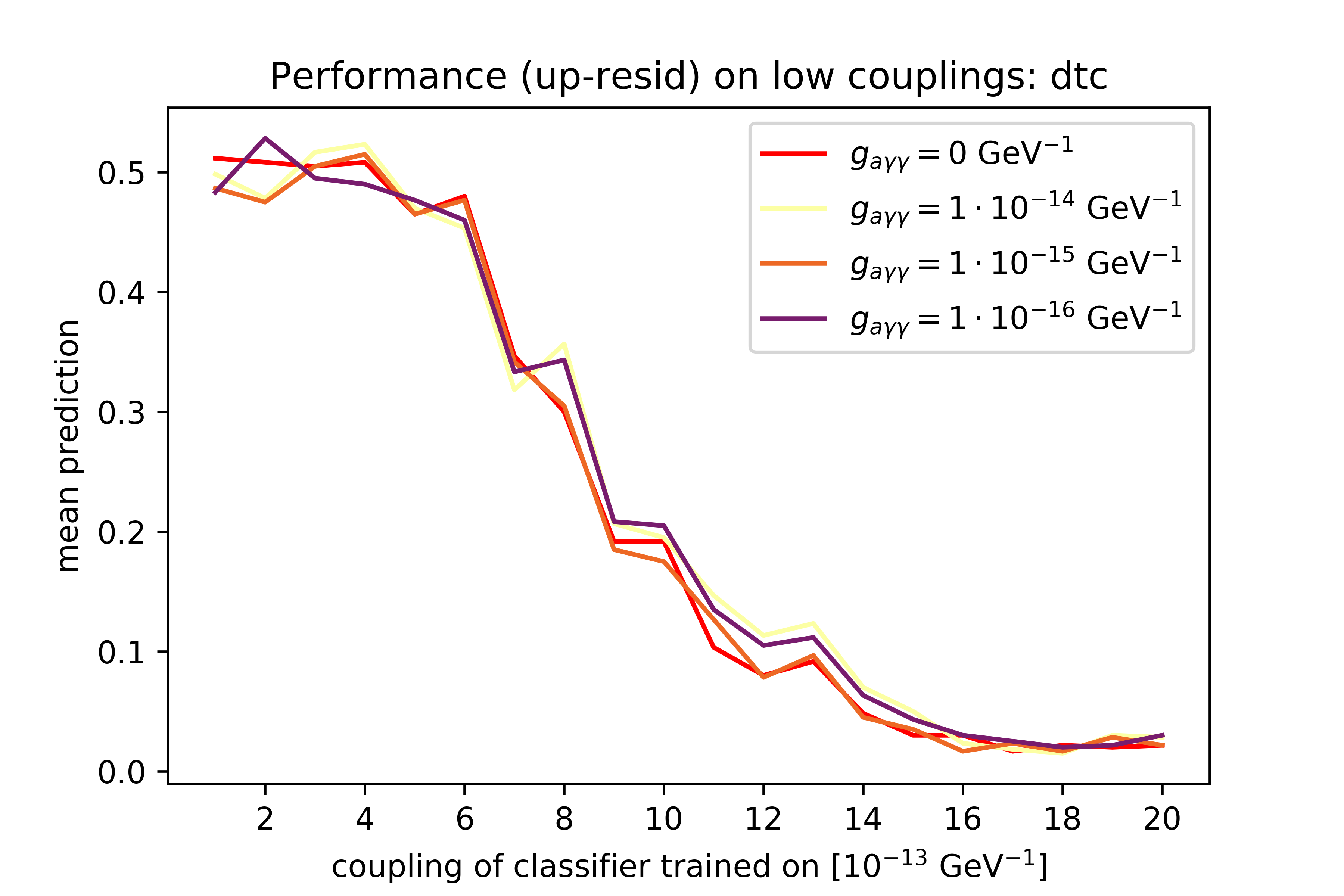}
\caption{As Figure~\ref{fig:ngc1275summary} but for very low values of $g_{\rm query}$.}
\label{fig:ngc1275low}
\end{figure}

\begin{figure}[H]
\hspace{-1cm}
\includegraphics[width=0.55\textwidth]{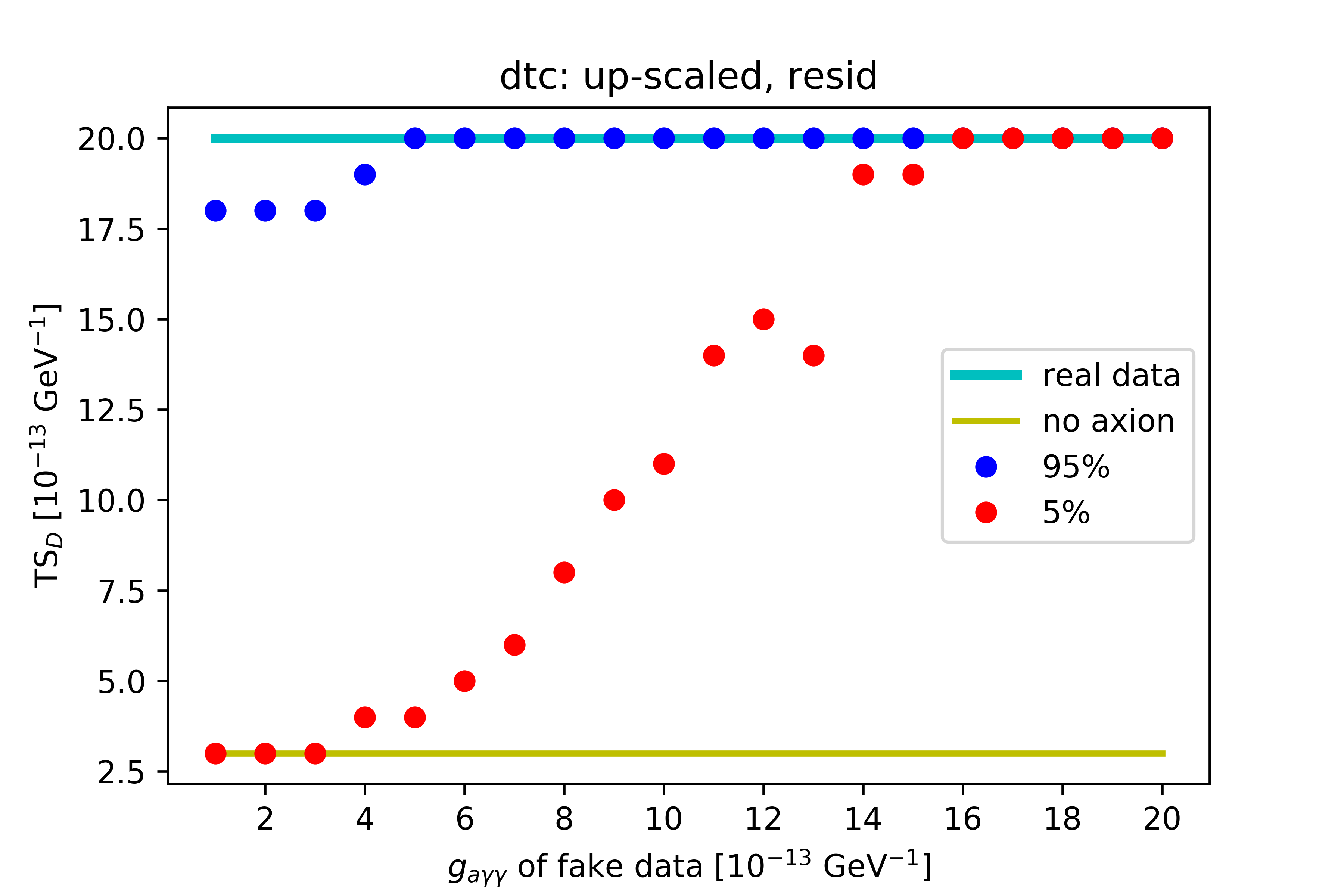}
\includegraphics[width=0.55\textwidth]{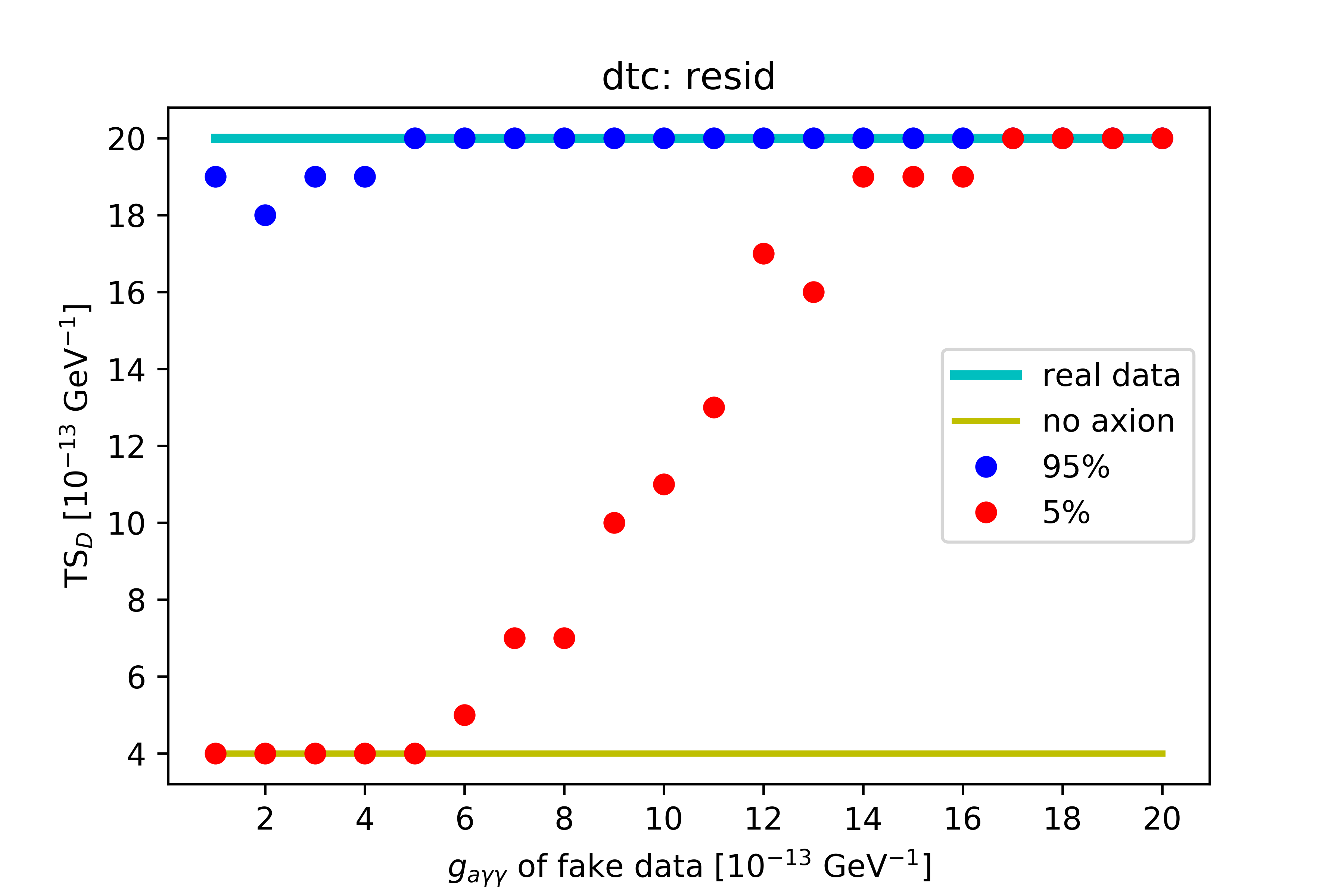}
\caption{Test statistic quartiles for the DTC classifier applied to up-scaled residuals (left) and for residuals (right) for NGC1275. We query the classifiers with simulated data with $g_{a \gamma \gamma}$ as shown on the $x$ axis. The $y$ axis shows the $5 \%$ (red)  and $95 \%$ (blue) percentile values of the test statistic defined above. The blue line is the test statistic of the real data and the yellow line is the test statistic of the no-ALP data.}
\label{fig:ngc1275dots}
\end{figure}

In this appendix we present a short overview of the performance we find for the central AGN of NGC1275 in the Perseus cluster. As the available data features significantly more counts, we can train our classifiers with less noisy data samples. In turn this leads to a very good performance of the classifiers. Figures~\ref{fig:ngc1275summary},~\ref{fig:ngc1275low}. and~\ref{fig:ngc1275dots} show examples of the performance we observe. However, we are unable to place bounds using this method as the real data is consistently classified as `axiony' due to anomalies in the observed spectra. The performance of our classifiers suggests that the constraining or discovering power of NGC1275 is very high, potentially reaching $g_{a \gamma \gamma} \sim 4 \times 10^{-13} \, {\rm GeV}^{-1}$ even for {\it Chandra} data taken without the High Energy Transmission Grating.

\begin{figure}[H]
\centering \includegraphics[width=0.55\textwidth]{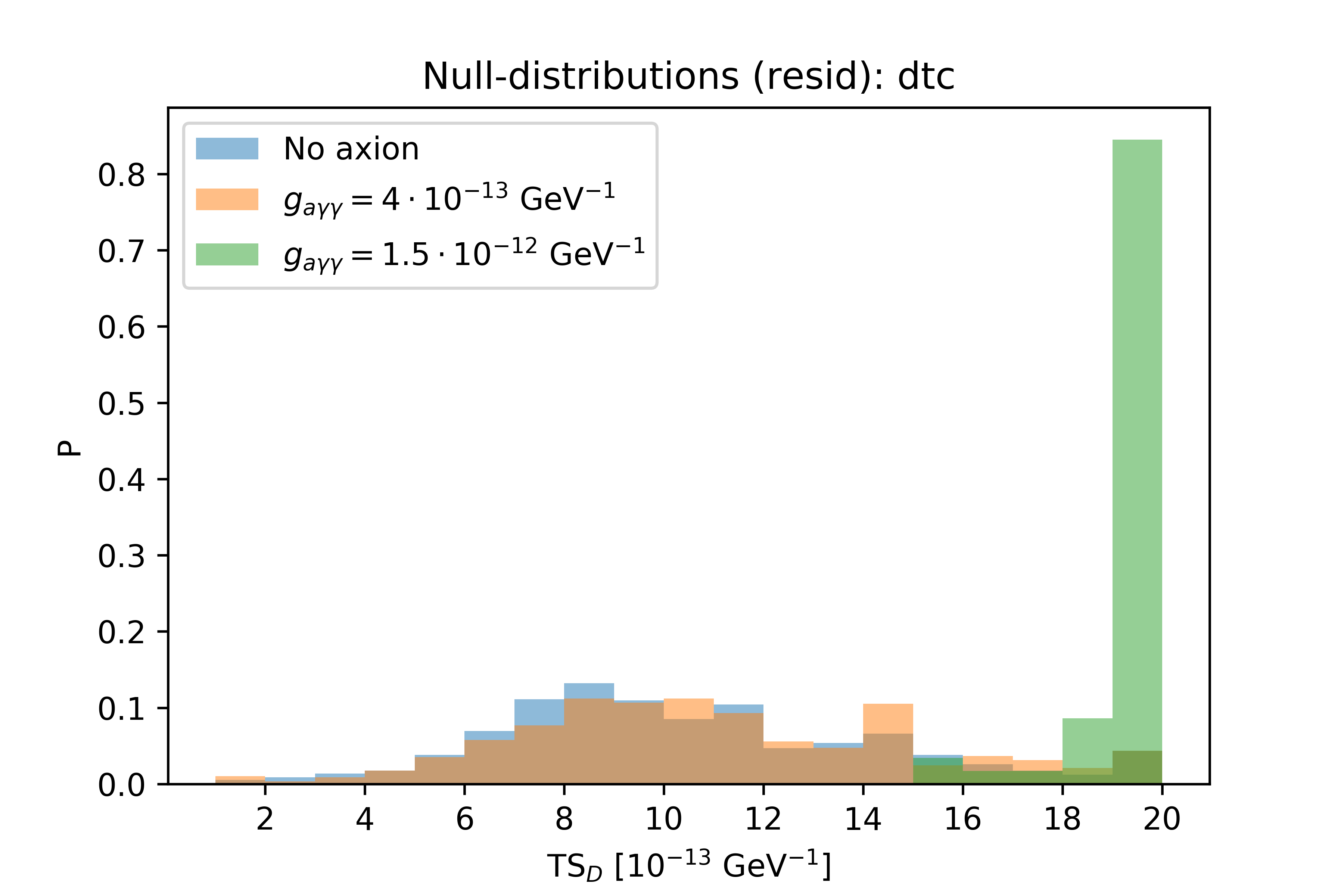}
\caption{Null distribution for the DTC classifier on residual data for NGC1275}
\label{fig:residual}
\end{figure}

\section{3D Magnetic field simulations}
\label{app:newmagneticfields}

For the analysis presented in the main body of the paper, we used magnetic fields simulated in one dimension only, along the line of sight to the source. This simplification is required so that simulating the quantity of training and test data required is computationally tractable. It is natural to ask whether a different choice of magnetic fields leads to significant changes on our classification.\footnote{We would like to thank the referee for highlighting this point.} We therefore also test the performance of our classifiers on spectra simulated with ALP-photon interconversion in the presence of full three-dimensional turbulent field simulations, as described in \cite{1603.06978,1312.3947}. Unlike the one-dimensional simulations used for the results presented above, this more sophisticated simulation does not have a discrete domain structure, and the magnetic field direction changes smoothly everywhere.

We take as an example the quasar B1256+281 located behind the Coma galaxy cluster. We use the same parameters for Coma's field as in our one-dimensional simulations, but simulate the turbulent field as follows:

\begin{enumerate}
\item Generate a random vector potential field in Fourier space according to $\langle | \tilde{\bf A}(k) |^2 \rangle \sim |k|^{-n}$. We allow $k$ to take values from $k_{\rm min} = \frac{2 \pi}{\Lambda_{\rm min}}$ to $k_{\rm max} = \frac{2 \pi}{\Lambda_{\rm max}}$. The phase of each component of $\tilde{\bf A}(k)$ is uniformly distributed between $0$ and $2 \pi$. We use a Kolmogorov power spectrum corresponding to $n = 17/3$, $\Lambda_{\rm min} = 2$ kpc and $\Lambda_{\rm min} = 34$ kpc \cite{1002.0594}. 

\item The Fourier space magnetic field is given by $\tilde{\bf B}(k) = i {\bf k} \times \tilde{\bf A}(k)$.

\item We find the real space field ${\bf B}({\bf x})$ by taking the Fourier transform of  $\tilde{\bf B}(k)$ and normalising the result to the correct amplitude based on distance from the cluster centre, as described in \cite{1704.05256}. We simulate ${\bf B}({\bf x})$ on a $2000^3$ grid with cell size $0.5$ kpc.

\item We then calculate the photon survival probability for five different sight-lines at a projected 240 kpc from the cluster centre, corresponding to the quasar's location.

\end{enumerate}

\begin{figure}
\begin{center}
\includegraphics[width=0.32\textwidth]{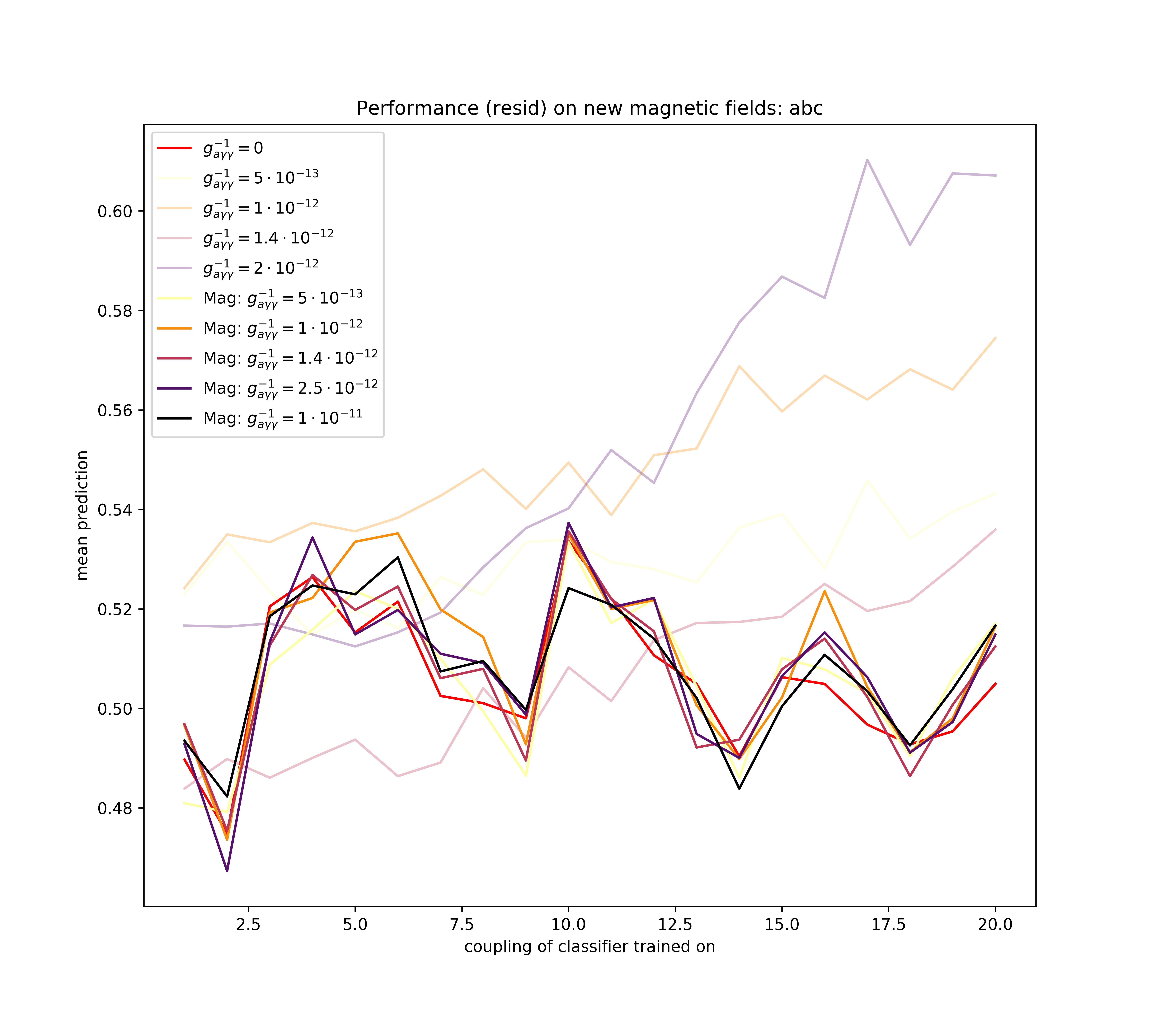}
\includegraphics[width=0.32\textwidth]{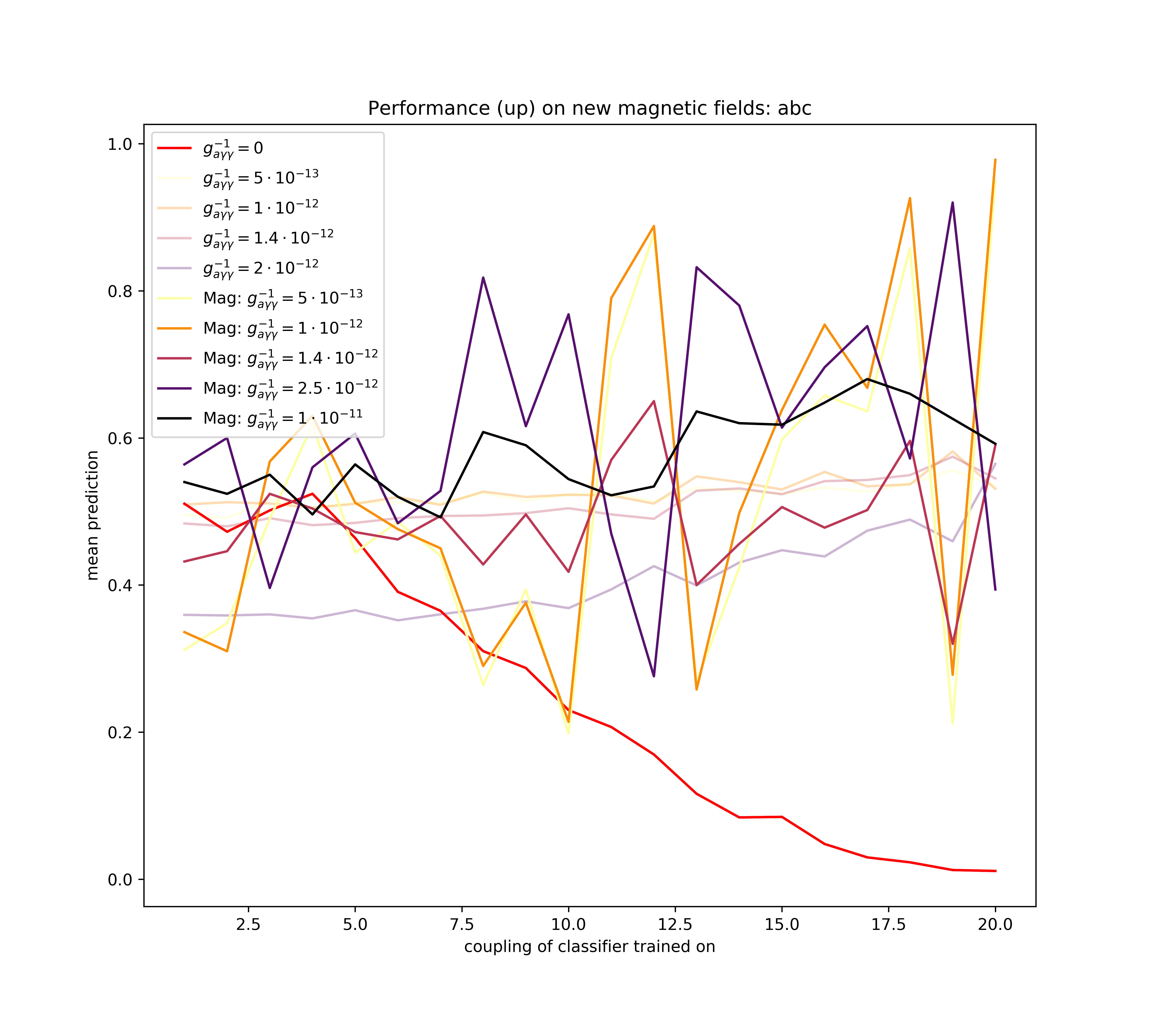}
\includegraphics[width=0.32\textwidth]{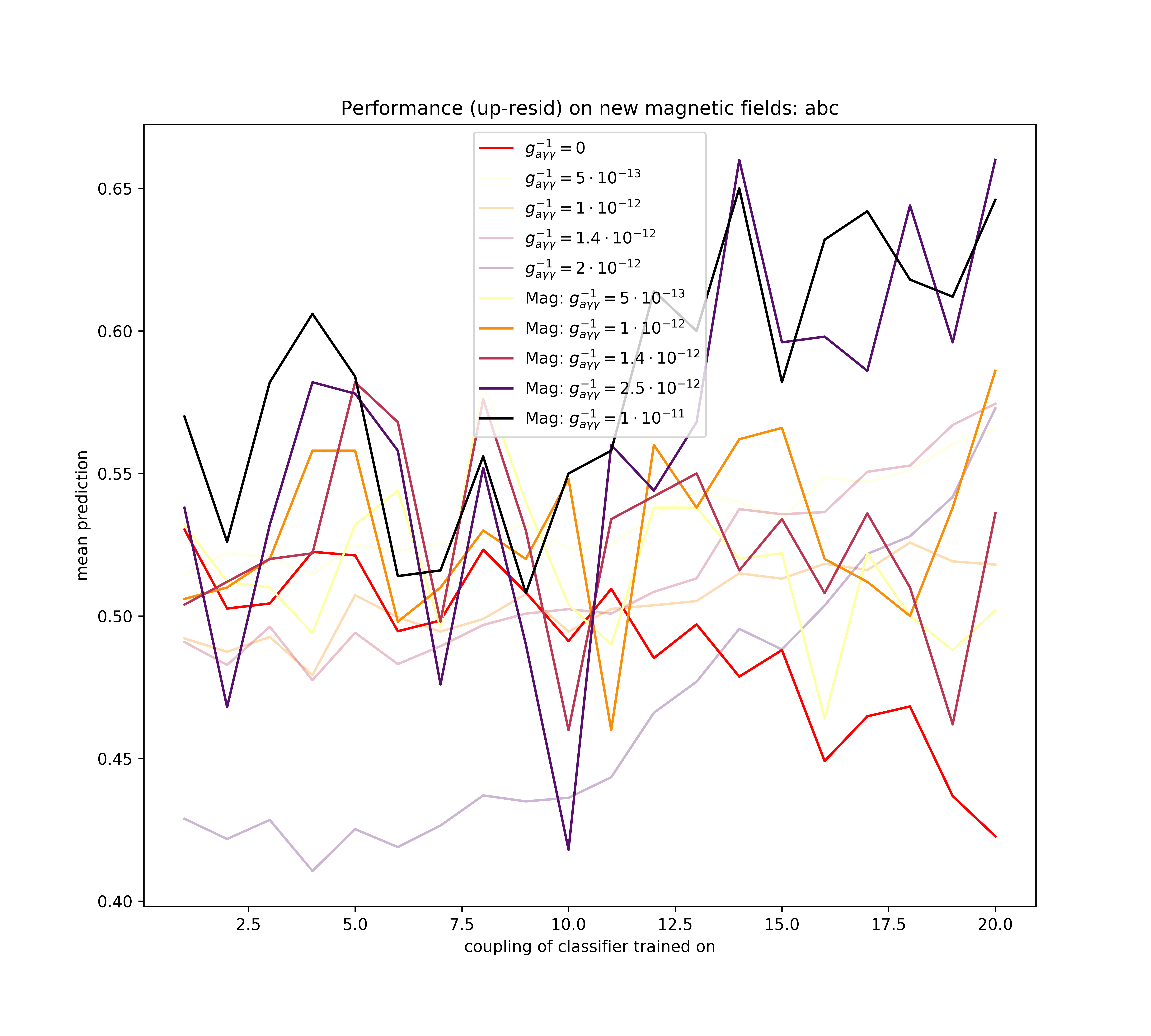}
\end{center}
\caption{{\bf Left:} Comparison of the mean performance of our ABC classifiers with old and new magnetic fields for residual data. {\bf Middle:} Comparison of the mean performance of our ABC classifiers with old and new magnetic fields for up-scaled data. {\bf Right:} Comparison of the mean performance of our ABC classifiers with old and new magnetic fields for up-scaled residual data. The new magnetic field data is shown by the bolder lines labelled `Mag'. The underlying source is the quasar SDSS J130001.48+275120.6 shining through the Coma cluster. The higher oscillatory behaviour in comparison to the previous magnetic field is typical for datasets which are based on few magnetic field realisations.\label{fig:newmagneticfield}}

\end{figure}

Associated to this magnetic field model we have generated five photon survival probabilities each for five different couplings in the range $g_{a\gamma\gamma}^{-1}=5\cdot 10^{-13}-1\cdot 10^{-11}.$
For each of these 25 survival probabilities we have generated 100 fake spectra as described in Section~\ref{sec:datasets}. We have then classified these new spectra with our classifiers which were trained with the previous datasets based on one-dimensional magnetic fields. The mean predictions found for the classifiers are compared with the previous predictions for spectra simulated with the one-dimensional fields, and three examples for up-scaled, residual, and up-scaled residuals on the AdaBoostClassifier are shown in Figure~\ref{fig:newmagneticfield}. We find very similar results for all classifiers. For the latter two, we see that in the regime where the classifiers are working at reasonably large couplings and for data with ALPs of large enough couplings, the mean performance is very similar. We cannot identify a large change due to this change in the magnetic field for these two data-products. As the number of magnetic field realisations is relatively low, we are not surprised by the higher oscillatory behaviour in the mean performance.

However, for the residual data we find that the ALP data is tracing the no-ALP mean prediction and not the ALP mean-predictions. The classifiers identify these spectra with ALPs as no-ALP. At this stage, we do not understand the underlying reason for this phenomena. Similarly to the unexpected behaviour of the up-scaled data on low couplings, a detailed investigation is beyond the scope of this article. For now, we simply consider these bounds as demonstrative of the potential of machine learning, rather than giving us true bounds on $g_{a \gamma \gamma}$.

\bibliography{MLrefs}{}
	\bibliographystyle{JHEP}

\end{document}